# Understanding Labor Market Discrimination Against Transgender People: Evidence from a Double List Experiment and a Survey[♠]

Billur Aksoy[♥]    Christopher S. Carpenter[♦,*]    Dario Sansone[♣]

**Abstract**

Using a US nationally representative sample and a double list experiment designed to elicit views free from social desirability bias, we find that anti-transgender labor market attitudes are significantly underreported. After correcting for this concealment, we report that 73 percent of people would be comfortable with a transgender manager and 74 percent support employment non-discrimination protection for transgender people. We also show that respondents severely underestimate the population level of support for transgender individuals in the workplace, and we find that labor market support for transgender people is significantly lower than support for gay, lesbian, and bisexual people. Our results provide timely evidence on workplace-related views toward transgender people and help us better understand employment discrimination against them.

[♠] Funding: The Vanderbilt LGBTQ+ Policy Lab. The list experiment and a pre-analysis plan are pre-registered on the American Economic Association's registry for randomized control trials (AEARCTR-0008820): https://doi.org/10.1257/rct.8820. We thank Brad Barber, Ian Chadd, Marcus Dillender, Keith Marzilli Ericson, Gilbert Gonzales, Oliver Hauser, Julian Jamison, Boon Han Koh, Max Lee, Tara McKay, David McKenzie, Sonia Oreffice, Connor Redpath, and Lisa Windsteiger for their helpful comments. We also thank seminar participants at AEA CSQIEP, CREST, LISER, Monash University, Universidad de Alicante, Universidad Autónoma de Madrid, University of Exeter, Utrecht University, Vanderbilt University, and conference participants at the COST Conference on LGBTQ+ Methodologies, the 56[th] Annual Canadian Economics Association Meetings, the 2022 World Economic Science Association Conference, and the UEA Discrimination and Diversity Workshop for their feedback. All errors are our own.

[♥] Rensselaer Polytechnic Institute. E-mail: aksoyb3@rpi.edu
[♦] Vanderbilt University, NBER, and IZA. E-mail: christopher.s.carpenter@vanderbilt.edu
[♣] University of Exeter and IZA. E-mail: d.sansone@exeter.ac.uk
[*] Corresponding author

# 1. Introduction

Very little is known about labor market discrimination against transgender people.[1] This is in sharp contrast to a substantial and growing literature on the employment experiences of sexual minority populations relative to heterosexual individuals (Klawitter 2015; Neumark 2018; Badgett, Carpenter, and Sansone 2021) and on attitudes toward sexual minorities in the workplace and support for employment non-discrimination protection on the basis of sexual orientation (Coffman, Coffman, and Ericson 2017). In this paper, we study Americans' views about transgender managers in the workplace, as well as their support for employment non-discrimination protection for transgender individuals using a representative online sample of the US population.

Understanding labor market views toward transgender people is important especially in the context of the 2020 US Supreme Court decision in *Bostock v. Clayton County* which ruled that transgender people are legally protected from discrimination in employment. Multiple recent studies using population data on transgender people have demonstrated that gender minorities have significantly worse economic outcomes than otherwise similar cisgender people, even though employment discrimination against transgender people is illegal (Badgett, Carpenter, and Sansone 2021; Carpenter, Eppink, and Gonzales 2020; Carpenter, Lee, and Nettuno 2022). Nevertheless, we do not have good economic data on how transgender people are treated by employers, co-workers, or the general public with respect to labor market outcomes. Understanding these attitudes is important as they could affect health outcomes and disparities (National Academies of Sciences Engineering and Medicine 2020) through minority stress – i.e., stress due to internalized homophobia and transphobia, anticipated rejection, constant efforts to hide one's identity, and actual experiences of discrimination and violence (Meyer, 1995). In addition, studying the level of support for employment protection is important for contextualizing its potential effectiveness and for improving relative outcomes for transgender people in the US. Furthermore, policymakers discussing proposed transgender-related policies may want to know whether voters support such policies, and employers or managers considering hiring and promoting transgender individuals may want to know if those individuals would be supported in the workplace.

The fact that we have a very limited understanding about attitudes toward transgender employment rights and transgender people in the workplace is problematic also because a nontrivial share of the population identifies as transgender. Recent Pew Research Center data indicated that 1.6 percent of adults identified as transgender in 2022; the rate among adults under age 30 was 5.1 percent (Brown 2022a). Moreover, 44 percent of adults reported knowing someone who is transgender. These survey-based estimates are likely lower bounds due to individuals' concerns about social stigma and potential discrimination.[2] Therefore, understanding views toward these populations is important as transgender individuals represent a substantial and growing minority.

---

[1] Transgender people are individuals whose sex assigned at birth as either male or female differs from their current gender identity or expression as a man, woman, both, or neither; cisgender people are those whose sex assigned at birth matches their current gender identity. Transgender individuals and other gender diverse populations are sometimes referred to as gender minorities. Gay men, lesbian women, bisexual, asexual and queer individuals are referred to as sexual minorities.

In this paper, we study views toward transgender people in the workplace and support for transgender-related employment non-discrimination rights using an online sample that is representative of the US population with respect to race, sex, and age. Eliciting views about transgender people in the workplace and about transgender employment rights may be susceptible to social desirability bias. For instance, such biases may exist because of the perception that expressing anything other than support for transgender people in the workplace could result in negative reprisals (due, for instance, to the recent rise of 'cancel culture'). This would result in an artificially high rate of stated support for transgender people in the workplace. We overcome these biases – and document their importance and magnitude – by being the first to study transgender-related labor market views using a list experiment technique. This technique has been widely used in psychology, sociology, political science, and economics to elicit sensitive views and attitudes free from social desirability bias.[3] In a list experiment, individuals are presented with a list of statements and asked to report how many of the statements in the list are true for them, but they are not asked whether each specific statement is true for them. In our list experiments, one group of respondents is presented with four statements and another group is presented with the same four statements plus an additional key statement of interest pertaining to their views about transgender people in the workplace (specifically, whether they would be comfortable having a transgender manager or whether they support employment non-discrimination protection for transgender people). Comparisons across lists allow us to back out an estimate of the true share of respondents who agree with each key statement of interest regarding transgender people in the workplace.

While the list experiment technique cannot identify *which* specific individuals agree with the key statements (because individuals only report the total number of statements within each list that are true for them as opposed to indicating whether each individual statement is true for them), it has the distinct advantage that we can credibly estimate population-level views toward transgender people in the workplace that are free from social desirability bias. Additionally, toward the end of our survey, we directly ask respondents about the key statements of interest (comfort with a transgender manager and support for employment non-discrimination protection for transgender people), which, when compared with the true share elicited through the list experiments, provides us with estimates of the magnitude of misreporting of attitudes regarding transgender people in the workplace. We can also use group characteristics to examine whether, for example, women on average are more or less supportive of transgender people in the workplace than men. Finally, as discussed in more detail in Section 3, we use a *double* list experiment to verify the robustness of our findings to using different non-key statements (Chuang et al. 2021) and to increase the precision of our estimates by minimizing the variance (Droitcour et al. 1991; Glynn 2013).

---

[2] These shares are also increasing over time. For example, the 2017 version of the Pew Research Center survey found that only 37 percent of adults knew a transgender person, and this rose to 42 percent in 2021 and 44 percent in 2022. Regarding transgender identification, BRFSS estimates for a large number of US states suggest that about 0.5 percent of adults identify as transgender (Badgett, Carpenter, and Sansone 2021), and the most recent Gallup survey indicates that around 0.7 percent of adults do not identify as cisgender (Jones 2022). Moreover, the Gallup data – like the Pew Research Center data – reveal strong generational differences in reporting a non-cisgender identity (Jones 2022).

[3] We provide an overview of the literature on list experiments in Section 2. In section 3, we discuss the assumptions of list experiments and key design considerations.

Comparing our double list experiment to the direct survey responses, we find that anti-transgender labor market sentiment is significantly underreported (by 6-7 percentage points), consistent with a strong role for social desirability bias. We also find that even after correcting for social desirability bias, 73 percent of people would be comfortable with a transgender manager at work and 74 percent support non-discrimination protection in employment for transgender people. Women, sexual minorities, and Democrats have significantly more positive views and show greater support than men, heterosexual individuals, and Republicans or Independents, respectively.

To complement the double list experiment, we then report the results from a descriptive survey. The survey allows us to compare views about transgender people in the workplace and about transgender employment non-discrimination rights in relation to views about lesbian, gay, and bisexual (LGB) people in the workplace and about LGB employment non-discrimination rights. In addition, our survey asked people about their general perceptions regarding the two statements of interest, i.e., their beliefs about the true population share of individuals who would be comfortable with transgender managers and who support employment non-discrimination protection for transgender people.

Looking at our survey data, we find that Americans show significantly higher support for LGB people in the workplace and for LGB employment non-discrimination rights relative to support for transgender people in the workplace and for transgender employment non-discrimination rights. Our survey data also demonstrate that respondents severely underestimate the true level of support for transgender people in workplace among the general population by 28 to 53 percent. This finding is especially notable given that beliefs about others' views on stigmatized behaviors are shown to impact individuals' *own* views and behaviors (Bursztyn, González, and Yanagizawa-Drott 2020). It may suggest that support for transgender people in the workplace could be increased by correcting biased beliefs.

Taken together our results provide timely evidence on labor market sentiment toward transgender people in the United States. Although anti-transgender sentiment is underreported, a sizable majority of American adults – nearly 3 in 4 – supports transgender people in the labor market, including in positions of workplace authority, and supports employment non-discrimination protection for transgender individuals. These findings are important given the documented positive effects of employment non-discrimination protections on wages for other minority groups (Donohue and Heckman 1991; Klawitter and Flatt 1998; Neumark and Stock 2006; Klawitter 2011; Delhommer 2020).

## 2. Literature review

Our study is related to a large economics literature on the drivers and impacts of discrimination in labor markets (Arrow 1973; Phelps 1972; Becker 1971; Bertrand and Duflo 2017; Neumark 2018). There is also a vast literature on discrimination based on social identity (such as race and gender) (Altonji and Blank 1999; Goldin and Rouse 2000; Bertrand and Mullainathan 2004; Lang and Spitzer 2020). Within this large body of literature, recent research has shown that LGBTQ+ individuals are subject to discrimination in formal markets such as labor and housing (for a review, see Badgett, Carpenter, and Sansone 2021) as well as in domains outside of these formal contexts such as with respect to prosocial behavior (B. Aksoy, Chadd, and Koh 2021).

A small economics literature on employment, earnings, and income for transgender people also has emerged, with most studies finding that transgender people have significantly worse economic outcomes than similarly situated cisgender people (Badgett, Carpenter, and Sansone 2021; Geijtenbeek and Plug 2018; Granberg, Andersson, and Ahmed 2020; Shannon 2022; Carpenter, Eppink, and Gonzales 2020). For example, the most recent evidence from nationally representative US data indicates that non-cisgender individuals have significantly lower employment rates and higher poverty rates than otherwise similar cisgender individuals (Carpenter, Lee, and Nettuno 2022). We contribute to this broad but relatively new body of literature by studying views about transgender managers in the workplace and support for employment non-discrimination protection for transgender individuals. The comparison of views toward transgender individuals relative to LGB individuals in the workplace also provides an important contribution to this literature.

As we examine comfort with having a transgender manager, our paper extends the literature examining the employment barriers (e.g., "glass ceilings") faced by women, racial minorities, and sexual minorities in accessing positions of leadership (Albrecht, Björklund, and Vroman 2003; Frank 2006; Giuliano, Levine, and Leonard 2009; Matsa and Miller 2011; C. G. Aksoy et al. 2019; Cullen and Perez-Truglia 2021). We are not aware of any other research that directly examines managerial or supervisory authority among transgender individuals. We also contribute to the growing literature on attitudes towards transgender individuals (Broockman and Kalla 2016; Taylor, Lewis, and Haider-Markel 2018; Luhur, Brown, and Flores 2019; McCarthy 2021; Lewis et al. 2022; Doan, Quadlin, and Powell 2022).

Our paper contributes to the literature on list experiments. Several studies in psychology, sociology, and political science have used list experiments to elicit sensitive views and attitudes, including in the context of sexual minority rights. For example, Lax, Phillips, and Stollwerk (2016) have used a list experiment to measure public support for same-sex marriage in the US, finding no evidence of social desirability bias regarding support for same-sex marriage or the inclusion of sexual minority status in employment non-discrimination laws. Other research in these fields has used the list experiment to examine social desirability bias in the context of: support for a female American President (Streb et al. 2008), support for a Jewish presidential candidate (Kane, Craig, and Wald 2004); racial discrimination (Kuklinski, Cobb, and Gilens 1997; Kuklinski et al. 1997); the prevalence of Atheists (Gervais and Najle 2018); and the prevalence of risky sexual behaviors among college students (LaBrie and Earleywine 2000).

Within economics, list experiments have been more limited, with some notable exceptions. For example, development economists have used this method to study sexual activity and reproductive behavior in Uganda (Jamison, Karlan, and Raffler 2013) as well as Cameroon and Cote d'Ivoire (Chuang et al. 2021). List experiments have also been used in economics to examine corruption in public procurement in Russia (Detkova, Tkachenko, and Yakovlev 2021), use of loan proceeds in Peru and the Philippines (Karlan and Zinman 2012), illegal migration rates in Ethiopia, Mexico, Morocco, and the Philippines (McKenzie and Siegel 2013), hiring discrimination against women in Egypt (Osman, Speer, and Weaver 2021), and intimate partner violence in Peru (Agüero and Frisancho 2022).

Our study is most closely related to Coffman, Coffman, and Ericson (2017) who conducted a list experiment in 2012 to study anti-LGB sentiment using an Amazon Mechanical Turk sample. They showed that the magnitude of anti-LGB sentiment is significantly understated. Our results offer an important complement to their findings as the first list experiment evidence on views about transgender managers in the workplace and employment non-discrimination protection for transgender people.

### 3. Data and methodology

#### 3.1 Experimental design

#### 3.1.1 List experiments

We use a list experiment technique (also called "item-count technique", "unmatched count", or "veiled approach") that was pioneered by Miller (1984).[4] As mentioned in the introduction, respondents are given a list of statements and asked to report how many statements (but not which specific ones) are true for them, thus providing an extra layer of anonymity and increasing privacy (Coutts and Jann 2011). Participants are either assigned to a treatment group or a control group. In the control group ("short list"), participants are given a list of statements and asked to indicate how many of those statements are true for them. In the treatment group ("long list"), participants are given the same list of statements plus a key statement of interest (in our context, a statement about views towards transgender individuals in the workplace).[5] The difference in means between the

---

[4] We decided to use list experiments instead of randomized response technique (where respondents use a private randomization device - e.g., flip a coin - to determine whether they answer either a sensitive or innocuous question) because randomized response technique is more difficult to implement online, subjects trust the randomized response technique less than the list experiment (Coutts and Jann 2011), and participants may not respond to the randomization device relied upon by the randomized response technique as instructed (John et al. 2018).

[5] The order of the statements is randomized at the individual level in both the short and long lists. This serves two goals. First, if we do not randomize the order of the key statements and list them as last, as done by many papers in

two lists gives us the estimated share of the population with the key attribute of interest. Table 1 presents one of the lists used in our study.

{Table 1 here}

To formally illustrate how we use the list experiment technique to estimate the share of the population with the key attribute of interest, we follow the standard estimation technique implemented in previous studies (Tsai 2019). Suppose that we have a sample of *n* participants. Let $T_i$ be the indicator variable equal to one if participant *i* sees the long list instead of the short list, and 0 otherwise. Let $S_i$ be the potential answer to the key statement by participant *i*, and let $R_{i,j}$ be the potential answer to the jth non-key statement by participant *i* (where j=4 in our application). Using the list in Table 1, $S_i = 1$ if participant *i* would be comfortable having a transgender manager at work, and 0 otherwise. Similarly, for example, $R_{i,3} = 1$ if participant *i* can fluently speak at least three languages, and 0 otherwise. Recall that we do not observe $S_i$ or $R_{i,j}$. Instead, we observe the total number of statements that are true for participant *i*: $Y_i = T_i S_i + R_i$ where $R_i = \sum_{j=1}^{4} R_{i,j}$. Under certain assumptions,[6] the difference in means estimator as presented below gives us the estimated share of the population with the key attribute (i.e., $E(S_i)$).

$$E(S_i) = \frac{\sum_{i=1}^{n} Y_i T_i}{\sum_{i=1}^{n} T_i} - \frac{\sum_{i=1}^{n} Y_i(1 - T_i)}{\sum_{i=1}^{n}(1 - T_i)} \tag{1}$$

To increase power and reduce variance, we extend this technique by using double list experiments (Droitcour et al. 1991; Glynn 2013). For each key statement, we have a set of two lists, (e.g., list A and list B) that are designed to be positively correlated. Each list contains four non-key statements. Half of the participants (randomly selected) see list A (a short list) and then list B with the key statement (a long list). The other half see list A with the key statement (a long list) and list B (a short list). We also randomized the order at the subject level such that some participants see list A first while others see list B first. The differences-in-means between short and long lists from both lists A and B are averaged, providing us the true share of the population with that key attribute. Formally, let $Y_i^A$ and $Y_i^B$ be the total number of items in list A and B, respectively, that are true for participant *i*, the estimated share of the population with the key attribute is given by $E^{DL}(S_i)$.

$$E^{DL}(S_i) = \left[ \left\{ \frac{\sum_{i=1}^{n} Y_i^A T_i}{\sum_{i=1}^{n} T_i} - \frac{\sum_{i=1}^{n} Y_i^A (1 - T_i)}{\sum_{i=1}^{n}(1 - T_i)} \right\} + \left\{ \frac{\sum_{i=1}^{n} Y_i^B (1 - T_i)}{\sum_{i=1}^{n}(1 - T_i)} - \frac{\sum_{i=1}^{n} Y_i^B T_i}{\sum_{i=1}^{n} T_i} \right\} \right] / 2 \tag{2}$$

Thanks to this extension, it is possible to obtain more precise estimates since all respondents provide information about the key statements, unlike the single list experiment in which only

---

this literature, we worry that seeing a transgender-related statement as last in all lists could draw extra attention to the key statements. Second, the order of the statements might also have an impact on subjects' answers. By randomizing the order, we eliminate any aggregate effect coming from the ordering of the statements.

[6] The list experiment technique relies on three assumptions: treatment randomization, no design effect, and no liar. We discuss these assumptions and provide evidence in support of them in Online Appendix A.

respondents seeing the long list provide such information. The double list method also allows us to verify the robustness of our findings to using different non-key statements (Chuang et al. 2021).

In this experiment, we test two key statements:

*Transgender manager*: "I would be comfortable having a transgender manager at work."

*Transgender employment non-discrimination protection*: "I think the law should prohibit employment discrimination against transgender individuals."

We use the double list experiment technique for both statements and thus we have a total of four lists: Lists 1A and 1B for the transgender manager key statement and Lists 2A and 2B for the transgender employment non-discrimination protection key statement.[7]

Following the recommendation in the literature (Blair and Imai 2012; Aronow et al. 2015), we also ask questions directly regarding the key statements to all participants toward the end of the survey. The direct questions provide baseline estimates of the share of population with the key attributes, and this allows us to estimate the size of the bias due to social desirability and misreporting of stigmatized attitudes.

### 3.1.2 Survey questionnaire

All subjects first participate in the list experiment section and then move to the survey.[8] Subjects are not allowed to skip any questions in the list experiments and are not allowed to go back and revise their answers at any point. However, subjects are always free to leave the study whenever they wish. The order of the questions in the survey section is the same for all respondents. In addition to the two questions (relating to the two key statements from the list experiments) asked directly in the survey, we collect standard demographic and socio-economic variables, and we ask additional direct questions to measure participants' views toward LGB individuals in the workplace.

Finally, at the very end of the survey, we also elicited participants' beliefs about the two key statements used in the list experiment.[9] Specifically, the participants were shown the following statements and asked to fill in the blank with their best guess:

---

[7] Although it is common practice in the literature not to randomize the order of the lists, we chose to incorporate some randomization into our design to control for potential order effects (here, we refer to the order of the lists, not the order of the statements within the list). We provide more explanation on this in Online Appendix A and show that we do not find any significant concerns for order effects.

[8] At the beginning of the experiment, respondents signed a consent form and were informed that the purpose of the study was to understand the demographic composition of the respondents and their views on certain economic, political, and social issues. The description of the study did not specifically mention transgender issues, as we did not want to prime respondents or obtain a self-selected sample.

[9] We chose not to incentivize these questions in order to keep the study simple and relatively quick. Although we acknowledge the usual drawbacks of using an unincentivized elicitation method, we think that these data provide novel and valuable insights about participant behavior.

"Out of every 100 people in the general US population, I think approximately _____ out of 100 would be comfortable with having a transgender manager at work."

"Out of every 100 people in the general US population, I think approximately _____ out of 100 would agree that the law should prohibit employment discrimination against transgender individuals."

The complete set of instructions and survey questions used for our study can be found in Online Appendix C.[10]

### 3.2. Key design considerations

The list experiment technique allows researchers to estimate the true share of the population with the key attribute by providing an extra layer of anonymity to their responses. As discussed in the introduction, by comparing the responses in the list experiment to direct survey questions, we can also estimate the size of the bias due to social desirability and misreporting of stigmatized attitudes. Social desirability bias might cause some respondents not to report their true sentiments honestly when asked directly. This usually happens when the respondents believe that their opinion runs counter to the perceived social norm. Ex-ante, the size of the bias is not clear: online surveys may elicit truthful answers since they are self-administered, completed in private, and anonymous (Holbrook and Krosnick 2010; Robertson et al. 2018). Thus, the magnitude of misreporting we document is likely to be a lower bound to what might occur in other surveys, since most surveys are not conducted with as much privacy and anonymity and thus people may be less prone to social desirability bias even when answering the question directly.

Importantly, it is not the case that increased reporting under the veil of the list experiment is simply mechanical. Previous research has shown that list experiments provide increased estimates of prevalence only for stigmatized views: there is no evidence of this technique leading to an increase in reporting of innocuous behaviors (Tsuchiya, Hirai, and Ono 2007; Coffman, Coffman, and Ericson 2017).[11]

While designing the list experiments and choosing the non-key statements, we followed best practices in the literature (Glynn 2013). For example, one should carefully determine how many non-key statements to include. The number of non-key statements should be neither too low (to avoid a ceiling effect, i.e., participants reporting that all statements are true for them, thus removing the privacy protection provided by the list experiment) nor too high (to avoid higher variance and measurement error due to respondents' inability to remember or focus on all statements in the list). After carefully examining previous studies, we decided on four non-key statements. In each of the lists, we included a statement that we expected to be true for most people (to avoid a floor effect, i.e., participants reporting zero items, thus also removing the privacy protection provided by the list experiment), another statement that we expected to be false for most people (to avoid a ceiling effect), and the remaining two non-key statements were chosen such that they are expected to be negatively correlated.[12] This approach has the additional advantage of decreasing variance and increasing power. High variance is often an issue because the key statement is aggregated with a number of non-key statements. To some extent, the additional variance is the cost of the higher perceived privacy protection (Glynn 2013). Therefore, list randomization often produces results that are too high in variance to be statistically significant, especially if the attribute, view, or behavior of interest has low prevalence (Karlan and Zinman 2012). Thus, a modal response of 2 out of 4 for the non-key statements is desirable. Finally, in order to increase power further in the double list, we designed the non-key statements in Lists A and B to be positively correlated.

---

[10] The survey includes additional LGBT-related questions which are being analyzed in a companion paper.
[11] For instance, Coffman, Coffman, and Ericson (2017) did not find any significant misreporting when the additional key statement in the longer list was the following: "It has rained once where I live in the last four days".

Following Chuang et al. (2021), in order to draw less attention to our key statements and increase the validity of our list experiment, some of the non-key statements in our lists are political in nature. Additionally, instead of asking the direct questions right after their corresponding lists, in line with previous studies (Lax, Phillips, and Stollwerk 2016; Chuang et al. 2021), we ask the direct questions after the demographic questions, and together with other questions on income, religiousness, and political affiliation. This order was chosen to limit the participant's focus on the transgender-related statements in the list experiments. Additionally, following Berinsky (2004), we do not provide a "don't know" option in the direct question since individuals who hold socially stigmatized opinions may hide their opinions behind a "don't know" response. Finally, Coffman, Coffman, and Ericson (2017) showed that list experiments work better when the stigmatized answer in the related direct question is a "no" instead of a "yes". Thus, we designed our key questions such that the socially stigmatized answer is always a "no".

### 3.3. Data collection and study sample

We coded the study using oTree (Chen, Schonger, and Wickens 2016) and conducted it on an online platform, Prolific, which has been used in many economics studies (Zmigrod, Rentfrow, and Robbins 2018; Schild et al. 2019; Isler, Maule, and Starmer 2018; Oreffice and Quintana-Domeque 2021). Available evidence indicates some important advantages of Prolific over Amazon Mechanical Turk for conducting research: Prolific participants are more diverse, less dishonest, pay more attention to study instructions, and produce higher quality data (Peer et al. 2017; Palan and Schitter 2018; Eyal et al. 2021; Gupta, Rigotti, and Wilson 2021).

We ran our experiment in late January 2022 using Prolific's representative sample of the US population with respect to race, sex, and age. A total of 1,806 participants completed the study.[13]

---

[12] We check for ceiling and floor effects and present findings in Figures B1-B2 in Online Appendix B, which confirms they are negligible in our experiment. Online Appendix A provides further quality checks on our list experiment.

[13] We ran our study in two waves. During the first wave, 301 participants successfully completed the survey. During the second wave, which was conducted exactly one week later, 1,505 participants successfully completed the survey. We implemented a minor change to the instructions for the list experiment between the first and the second wave. Instructions can be found in Online Appendix C. Panels A and B of Table B3 in our Online Appendix report the

Participants never disclose any identifying information, and the survey is completely anonymous. The attrition rate was very low: a total of 36 participants started the study but did not complete it. Out of those 36, 25 exited the study before seeing the first list experiment. We only use the data of participants who completed the entire study. In addition, we included three attention check questions. Less than 1 percent (n=15) of the participants failed one out of the three attention checks. No participant failed two or more attention checks. Thus, we include all participants in our analysis. The study took about 7 minutes on average to complete, and subjects who successfully completed the study received $1.30 on average which corresponds to $10.40/hour.[14]

{Table 2}

In Table 2, we present summary statistics of our Prolific participants.[15] Comparing our sample to official population estimates from the Census and the American Community Survey (U.S. Census 2021; Ruggles et al. 2022), our sample appears representative not only based on age, ethnicity, and sex – as expected given the sampling methodology – but also with respect to income, marital status, employment status, and urbanicity. Our sample is similarly likely to be Republican but is more likely to be Democrat and less likely to be Independent, and our sample is also more educated than the general US population (U.S. Census 2021; GSS 2021). In terms of region, although we have slightly more people from the Northeast and less from the West, overall, the regional distribution is comparable to the US population.

In addition to our Prolific sample, we provide supplemental descriptive evidence from the American National Election Survey (ANES). The ANES is a large nationally representative survey of US adults that is widely used in political science and economics research (Morisi, Jost, and Singh 2019; Fouka and Tabellini 2022). We use publicly available microdata from the ANES 2020 Time Series Study.[16] We use ANES for two main purposes. First, these data include a 'feeling thermometer' type of question where respondents were asked to rate their feelings toward a variety of groups, including transgender individuals.[17] Below, when we investigate group-specific heterogeneity views about transgender people in the workplace (e.g., whether women report more positive views than men), we use the ANES patterns as a source of comparison and confirmation. Second, the ANES includes survey items that closely align with the questions we asked our Prolific

---

responses in the list experiments with and without pilot data and show that this minor change in the instructions did not have an impact on the reported views in the list experiment. Thus, we combine both data sets and report our findings using all 1,806 participants.

[14] We check the robustness of our findings by excluding participants who completed the study very quickly or very slowly (as measured by the top and bottom five percent of the study completion time distribution). Our main findings are robust, and the details are discussed in Online Appendix A.

[15] Tables B1-B2 in the Online Appendix report sample sizes based on sex at birth, gender identity, and sexual orientation.

[16] ANES 2020 data were collected in two waves: shortly before (between August 18, 2020 and November 3, 2020) and shortly after (between November 8, 2020 and January 4, 2021) the 2020 US Presidential Election.

[17] Specifically, the 2020 ANES asked respondents "How would you rate *transgender individuals*?" It also asked respondents "How would you rate *gay men and lesbians*?" Respondents were asked to provide a number between 0 and 100, with higher numbers indicating more positive views.

respondents, such as support for non-discrimination protection on the basis of sexual orientation.[18] As we explain below, the nationally representative ANES returns very similar patterns on questions that are common to both datasets, further suggesting that our Prolific sample is also likely to be representative of the US population.

## 4. Results

In this section, we first present our findings from the list experiment. We then report heterogeneity in workplace-related views toward transgender people based on participants' sex, sexual orientation, and political affiliation. Next, we examine participants' beliefs regarding other people's views towards transgender individuals in the workplace. After that, we describe results from the survey which compare views regarding lesbian, gay, and bisexual managers, and support for employment non-discrimination rights for sexual minorities to those for transgender managers and support for employment non-discrimination rights for transgender individuals, respectively.

### 4.1 Views towards transgender individuals in the labor market

First, we present our findings from the double list experiments and compare our data to the direct questions. The first two bars of Figure 1 present the proportion of our participants who are comfortable having a transgender manager at work (*Transgender Manager*) and the latter two bars present the proportion of participants who agree that the law should prohibit employment discrimination against transgender individuals (*Trans Employment Non-Discrim*). To estimate the true share of the population with the key attribute using the list experiments, we first take the difference in means between the long and the short lists for each key statement, separately for Lists A and B.[19] We then take the average of these two estimates. This average gives us the estimated proportion using the double list method which is presented as *Double List* in the figure. The *Direct Question* bars in Figure 1 are the shares of the population who report comfort with a transgender manager or support for employment non-discrimination protection for transgender people, respectively, that we estimate using the answers to the direct questions in the survey.

{Figure 1}

Looking at the first two bars of Figure 1, we find that discomfort with having a transgender manager in the workplace is significantly underreported. When asked directly, 80.1 percent of our participants say they would be comfortable having a transgender manager at work. However, when asked indirectly (i.e., using the double list experiment method), we find that the share of participants who would be comfortable with a transgender manager at work is only 73 percent, significantly lower than the estimates from the direct question.

---

[18] Specifically, the 2020 ANES asked respondents "Do you favor or oppose laws to protect gays and lesbians against job discrimination?" The ANES did not ask about support for non-discrimination protection for transgender people.

[19] Standard errors have been computed following (Glynn 2013): because estimation is accomplished by taking the difference in mean responses between two independent sets of respondents, the variance of the estimator can be calculated with the standard large-sample formula for a difference in means, and confidence intervals can be computed in the usual fashion. Furthermore, our estimates and standard errors reported in Figure 1 and Table B3 do not change when using the Stata command *kict ls* (Tsai 2019) performing least squares estimation specifically for a double list experiment. We also check the robustness of our findings by adjusting the standard errors for age, sex, and race stratification. Our main findings are robust, as shown in Panel C of Table B3.

These findings are similar when we look at the views towards employment non-discrimination protection for transgender individuals, which are presented in the latter two bars of Figure 1. When we directly ask participants whether they think that the law should prohibit employment discrimination against transgender individuals, 79.5 percent of them say yes. However, looking at our double list experiment, the estimated true percentage of participants who agree with this statement is 73.7 percent, which is significantly lower.

Overall, the percentage of the participants who are comfortable having a transgender manager at work and those who agree that the law should prohibit employment discrimination against transgender individuals decreases by 8.9 percent and 7.3 percent, respectively, when participants are provided an extra layer of privacy thanks to our double list experiment. This social desirability bias that we document in the context of transgender labor market attitudes is comparable in magnitude to Coffman, Coffman, and Ericson (2017) where they investigate sentiments towards lesbian, gay, and bisexual individuals in various contexts using a single list experiment.

Although we focus on the double list method when discussing our data since it gives us the highest precision, we also present our findings using the individual lists in Online Appendix Figure B3 and Panel A of Table B3 which show that our results are robust to using either list. Indeed, for both key statements, the difference between the estimate in List A and the one in List B is statistically indistinguishable from zero. These statistics confirm that our main results are robust across lists and are not driven by the choice of the non-key statements (Chuang et al. 2021).

Our findings using direct questions are broadly in line with previous estimates using similar questions. A 2016 survey reported 71.2 percent of respondents agreeing that "Congress should pass laws to protect transgender people from employment discrimination" (Flores, Miller, and Tadlock 2018) and a 2017 US representative survey reported 72.7 percent of the participants agreeing that transgender people should be protected from discrimination by the government (Luhur, Brown, and Flores 2019).[20] Finally, our results are also in line with a 2017 US representative sample vignette study that found 75 percent of Americans supporting employment non-discrimination protection for transgender individuals (Doan, Quadlin, and Powell 2022).

Next, we estimate the true population size with our two key attributes using a regression analysis. Since we used two lists for each key statement, we estimate the following regression model separately for each list and each key statement using OLS:

---

[20] There is a 5-8 percentage point difference when comparing our direct question results to these studies. This difference could be due to differences in the wording of the question and/or differences in the timing of the surveys, as the attitudes towards LGBT individuals have improved significantly over time (Gallup 2022).

$$y_i = \beta_0 + \beta_1 T_i + \beta_2 X_i + u_i$$

where $T_i$ is an indicator variable that takes the value of 1 if the list was long (i.e., with the key statement) or 0 if the list was short, and $X_i$ is the vector of control variables that includes state fixed effects, demographic controls (subject's age, sex at birth, race, sexual orientation, and sexual attraction), socio-economic controls (subject's education level, employment status, income, current political affiliation, and current religious affiliation), beliefs about general level of support for the key statements (i.e., support for transgender managers or employment non-discrimination protection for transgender individuals), and additional controls (whether at least one child less than 18 years of age lives in the subject's household, number of people living in the subject's household, marital status, and urbanicity). Thus, $\widehat{\beta_1}$ gives us the estimated true population size with the key attribute which is presented in Table 3. Panel A presents the estimated share of the participants who would be comfortable with a transgender manager at work and Panel B presents the estimated share of the participants who agree that the law should prohibit employment discrimination against transgender individuals.

{Table 3}

Columns 1 and 5 show the estimated share of the population without any controls. Thus, these estimated shares are the same as those presented in Table B3 Columns 1 and 2 of Panel A. Next, we find that our results are robust to the inclusion of control variables. As we add more controls, the estimated shares get slightly smaller for three out of four estimates. For only one of the estimates, the coefficient increases by a maximum of 1.1 percentage points. All of these provide strong support for findings discussed above in Figure 1 and Table B3.

Since we employed a double list experiment, we can take the average of the estimates from Lists A and B. Taking the average of the coefficients from our most conversative estimates (columns 4 and 8), we find that 71.9 percent of the participants would be comfortable with having a transgender manager at work and 74 percent of the participants agree that the law should prohibit employment discrimination against transgender individuals. These estimated proportions are significantly lower than the estimates obtained by using direct questions (*p-value < 0.001* and *p-value = 0.005*, respectively), further confirming the presence of social desirability bias.

To summarize, we show that a sizable majority of adults in the US supports transgender people in the labor market, including in positions of workplace authority. Almost three-fourths of individuals are comfortable with transgender individuals in positions of leadership in the workplace and support laws prohibiting employment discrimination against transgender individuals. However, we also show that many participants do not truthfully report their views regarding transgender individuals in the workplace when asked directly. This could be due to social desirability bias where some individuals may not feel comfortable expressing their actual sentiments on a socially sensitive topic. These findings imply that research conducted using only survey measures of views towards transgender individuals in the workplace may paint a more optimistic picture of the situation in the US than the reality.

**4.2 Perceptions about general views**

Next, we aim to understand what our participants think about the views of the general US population toward workplace issues related to transgender individuals. To do this, we elicited participants' beliefs about the two key statements used in the list experiment. More specifically, we asked participants' perceptions about views of the general US population towards transgender managers and employment non-discrimination protection for transgender individuals. Figure 2 presents these perceptions regarding comfort with having a transgender manager (Panel A) and support for employment non-discrimination protection for transgender individuals (Panel B).

{Figure 2}

Figure 2 presents two interesting take-away points. First, although the true proportion of our participants who are comfortable having a transgender manager at work is 73 percent, our participants guess on average that only 47.7 percent of the general US population would be comfortable with a transgender manager. That is, respondents underestimate the true level of comfort with a transgender manager by 25.3 percentage points (53 percent of the average guess). Similarly, although we estimated that 73.7 percent of our participants agree that the law should prohibit employment discrimination against transgender individuals, on average they think that only 57.4 percent of the general US population supports laws that prohibit employment discrimination – an underestimate of about 16.3 percentage points (28 percent of the average guess).

Second, our participants think that the general US population is more likely to support laws that prohibit employment non-discrimination than to be comfortable with a transgender manager (57.4 percent versus 47.7 percent, *p-value < 0.001*). This is an especially interesting finding given that we do not see a difference when we compare the estimated true proportions using the double list experiments in Figure 1 (73.7 percent versus 73 percent, *p-value = 0.812*).

We also study these beliefs separately for those who personally agree with the key statement when asked directly versus those who do not. These findings are presented in Figures B4 and B5. Both figures reveal that, perhaps not surprisingly, there is a positive correlation between individuals' own views and their beliefs (Spearman's Correlation coefficients are 0.34, *p-value < 0.001*, and 0.24, *p-value < 0.001* for transgender manager and transgender employment non-discrimination rights, respectively). In other words, people who *disagree* with the key statements (i.e., who state they would not be comfortable having a transgender manager or who do not support non-discrimination protection in employment for transgender individuals) guess lower levels of support from the general population than people who agree with the key statements.[21]

**4.3 Heterogeneity analysis**

In this section we study our main research questions by doing subgroup analyses. More specifically, we compare differences in means in the double list experiments and the direct questions across subgroups based on sex, sexual orientation, and political affiliation.[22] Results are presented in Figures 3 and 4.

{Figure 3}

{Figure 4}

First, we compare women's views to those of men's views (Panels A in Figures 3 and 4, as well as Table B4). Women have significantly more positive views about transgender individuals and show higher levels of support for employment non-discrimination laws relative to men. This is true for estimates using both the double list experiments and the direct questions. We find a similar gender difference using the nationally representative ANES data where women (relative to men) report significantly more positive feelings toward transgender individuals ($p\text{-value} < 0.001$). Furthermore, we find that both men and women misreport their true views, although the difference is not significant for men for the employment non-discrimination protection statement.

Second, we compare views by sexual orientation (Panels B in Figures 3 and 4, as well as Table B5).[23] We find that non-heterosexual individuals hold significantly more positive views than heterosexual individuals regarding transgender people in the workplace. However, the share of non-heterosexual individuals comfortable having a transgender manager (Panel B Figure 3) is higher than the associated share supporting employment non-discrimination protection for transgender individuals (Panel B Figure 4), and the difference in the level of support when compared to heterosexual individuals is smaller for the employment non-discrimination outcome in Figure 4 than for having a transgender manager in Figure 3. Moreover, looking at Panel B of Figure 3, we find that heterosexual individuals are significantly more likely to underreport the stigmatized view when asked about their comfort with having a transgender manager relative to non-heterosexual individuals, and this difference is substantial – more than 11 percentage points – and statistically significant at the five percent level (as indicated in Table B5). In fact, we do not find any significant evidence of misreporting by non-heterosexual individuals regarding their comfort with having a transgender manager: their views are similar across both elicitation methods. Looking at Panel B of Figure 4, we find that both heterosexual and non-heterosexual individuals misreport their true views about non-discrimination protection, and the misreporting is marginally significant for non-heterosexual individuals.

---

[21] There are several potential explanations. First, we know from the extensive research on social norms that individuals' own beliefs and actions tend to adhere to social norms (Bicchieri 2002). These beliefs may be indicative of individuals' perceived social norms on these sensitive issues and thus the positive correlation between individual views and the beliefs would be in line with this research. Second, this positive correlation may be due to a false-consensus effect, which is a cognitive bias that causes people to overestimate how much others are like them. However, it is interesting to note that, even among those comfortable with a transgender manager or who support employment non-discrimination protection for transgender individuals (Panels A in Figures B4 and B5), the average perceived levels of support among the US population are significantly lower than the ones estimated from the double list experiments in Figure 1. Finally, we also acknowledge it could be the case that, ex-post, people simply misreport their true beliefs to justify their (dis)agreement with those statements. Future research can shed more light on how these beliefs might interact with participants' own behavior.

[22] Following our pre-analysis plan, we also conduct subgroup analyses by race (Table B7), age (Table B8), sexual attraction (Table B9), socio-economic status (Tables B10-B13), religious affiliation (Tables B14-B15), and geographical location (Table B16). We do not find significant differences in support for transgender people in the workplace associated with race, income, or employment status. We do find that support for transgender people in the workplace is significantly higher among younger individuals, those who are not exclusively attracted to a different sex, and non-religious people.

[23] We classified those who answered yes to "Are you heterosexual/straight?" as heterosexual; and those who answered no as non-heterosexual.

Lastly, we also compare views across political affiliations. Results are presented in Panel C of Figures 3 and 4 (and Table B6). Several insights emerge. First, in both figures, Democrats' views regarding transgender individuals in the workplace are more positive than Independents' views, which are themselves more positive than Republicans' views using both elicitation methods. This political divide we observe in our dataset is consistent with the political divide in general acceptance of transgender individuals shown by a 2021 Pew Research Center survey (Brown 2022b). Similarly, it is consistent with the nationally representative ANES data where we find that Democrats report significantly more positive feelings towards transgender individuals relative to Independents (*p-value < 0.001*), who also report significantly more positive feelings compared to Republicans (*p-value < 0.001*). Second, we find significant underreporting of the stigmatized view about discomfort with having a transgender manager for all three groups. In contrast, when it comes to support for employment non-discrimination protection, we only see significant misreporting by Independents. Meanwhile, the estimated support for employment non-discrimination for both Republicans and Democrats is similar across the two elicitation methods. In line with this, the only significant difference in the extent of misreporting arises when we compare Democrats to Independents (Table B6).

Next, we present regression results where we control for sex, race, age, sexual orientation, sexual attraction, political affiliation, household income, employment status, religious affiliation, region and beliefs. We estimate the heterogenous effects of these independent variables using an estimation method specifically designed for double list experiments by Tsai (2019).[24] This method estimates Equation 2 using a linear least-squares estimation method while controlling for independent variables as well as interacting them with the treatment variable. These results are presented in Table 4 separately for the key statement about having a transgender manager (Column 1) and the key statement regarding employment non-discrimination protection (Column 2).

---

[24] We use the Stata command *kict ls* (Tsai 2019).

{Table 4}

Overall, the heterogeneity findings presented above are in line with these estimation results. Women and non-heterosexual individuals hold more positive views regarding transgender individuals, although the coefficient estimates are not statistically significant for the employment non-discrimination protection statement. Table 4 confirms our results regarding how one's political party affiliation correlates with their views towards transgender managers and employment non-discrimination protection. In line with our findings discussed in Section 4.2, there is a positive correlation between participants' own views and their beliefs.[25] Table 4 also reveals that participants with less than a Bachelor's degree have significantly less positive views towards transgender managers. We do not see a significant difference in views across different age groups, religious affiliations, income levels, employment status, or regions.

Finally, although not specified in our pre-analysis plan, we also report evidence on heterogeneity in support for transgender individuals in the workplace related to prior managerial experience.[26] Individuals with such experience might plausibly have more information about managerial duties and responsibilities, and they are also more likely to be in positions that must comply with new non-discrimination regulations post-*Bostock*. We find that support for transgender individuals in the workplace is higher among individuals *without* managerial experience (Table B13). Moreover, the difference between the double list estimates and the answers to the direct question on comfort with a transgender manager is larger among those with managerial experience (*p-value = 0.101*); i.e., individuals with managerial experience misreport more than individuals without managerial experience.[27] These patterns may indicate that targeted managerial-focused interventions may be needed to ensure the equal treatment of transgender people in the workplace.

## 4.4 Comparison of workplace-related views toward transgender individuals relative to LGB individuals

So far, we have focused our analysis on views regarding transgender managers and support for employment non-discrimination protection for transgender people. It is also interesting to examine how these views compare relative to views regarding lesbian, gay, and bisexual individuals in these same contexts. As described in Section 3, in the survey we asked questions that allow us to examine these differences directly. Results are presented in Figure 5.

---

[25] These correlations are also clear from the raw differences in means by beliefs (Table B17). In particular, the difference between the estimated level of support for employment discrimination protection from the double list experiment and from the direct question is significantly larger among those who believed that most Americans would support this policy. That is, we find higher social desirability bias among respondents who believe most Americans would support employment discrimination protection for transgender individuals.

[26] We did not ask about managerial experience in our survey, but Prolific collects that information for a majority of the sample, and we use that information here.

[27] These patterns with respect to prior managerial experience are especially interesting given that such experience is positively correlated with education, and we see the opposite pattern for education: individuals without a bachelor's degree have significantly less comfort with a transgender manager than individuals with a bachelor's degree or higher. Together, these patterns suggest that there is something unique about managerial experience that is related to negative views toward transgender people in the labor market.

{Figure 5}

We find that support for transgender managers in the workplace is significantly lower than support for lesbian, gay, and bisexual managers (see first two bars of Figure 5). Participants are 9.6 percentage points less likely to report being comfortable having a transgender manager relative to an openly lesbian, gay, or bisexual manager. Looking at support for employment non-discrimination protection (the latter two bars of Figure 5), again, we see that participants are less likely to support such laws when those laws are designed to protect transgender individuals as opposed to lesbian, gay, and bisexual individuals. This pattern is further supported by the nationally representative ANES data indicating that feelings toward lesbian women and gay men are significantly more positive than feelings toward transgender individuals (*p-value < 0.001*).[28] The pattern is also consistent with previous studies measuring attitudes towards sexual and gender minorities (Lewis et al. 2017; Flores, Miller, and Tadlock 2018; Lewis et al. 2022).[29]

## 5. Conclusion

We report the results of a double list experiment and a survey designed to provide timely information on Americans' views toward transgender people in the workplace and support for transgender employment non-discrimination rights. As sexual and gender minorities are newly protected by federal employment non-discrimination protections as recently as Summer 2020, we sought to gauge workplace-related sentiment toward gender minorities using an elicitation method that removes social desirability biases which might artificially inflate support for transgender people in the workplace and transgender employment non-discrimination rights.

Our double list experiment yielded three key findings. First, anti-transgender labor market sentiment in our representative online sample was significantly underreported, consistent with the presence of social desirability bias and pressure to report comfort with transgender managers and support for transgender employment non-discrimination protections. Second, despite the presence of significant underreporting of anti-transgender sentiment, overall levels of true comfort with having a transgender manager at work and support for employment non-discrimination protection for transgender people were well over 70 percent. Thus, a sizable majority of adults in the US support transgender people in the workplace and transgender employment non-discrimination rights. Third, this support varied across demographic groups, with more support among women, sexual minorities, and Democrats.

---

[28] For reference, ANES data indicate that Americans have more positive feelings toward Jewish people and Black people than toward transgender individuals. Americans also have similar feelings towards Muslim and transgender individuals, while their feelings toward transgender people are more positive than their feelings toward feminists and individuals who participate in the Black Lives Matter movement.

[29] Notably, the share of our Prolific respondents who support employment non-discrimination for sexual minorities (84.9 percent) is very similar to the share of nationally representative ANES respondents who favor laws to protect gay men and lesbian women against job discrimination (86.6 percent). Moreover, the shares of our respondents who support LGB managers (89.7 percent) and LGB non-discrimination (84.9 percent) are comparable to Coffman, Coffman, and Ericson (2017) where 83.8 percent of their Mechanical Turk participants indicated that they would be happy to have a lesbian, gay, or bisexual manager at work and 85.6 percent said that they believe it should be illegal to discriminate in hiring based on someone's sexual orientation. Thus, our data on support for LGB people in the workplace are in line with previous well-designed surveys, including the nationally representative ANES that was fielded less than 24 months prior to our experiment.

Our survey yielded additional insights on views toward transgender people in the labor market in the United States. We found that people severely underestimate the level of comfort with having a transgender manager at work and the level of support for employment non-discrimination protection for transgender people. We also found that survey respondents reported more support for lesbian, gay, or bisexual people in the workplace and employment non-discrimination rights for lesbian, gay, or bisexual individuals than for transgender people in the workplace and for transgender employment non-discrimination rights, respectively.

Our results are highly relevant for policy. Indeed, they show large popular support behind the 2020 Supreme Court ruling in *Bostock v. Clayton County* banning employment discrimination against transgender people. They also emphasize the importance of accounting for social pressure when measuring support for sensitive policies, since people may misreport their true beliefs: people's actual views are the ones that will guide their voting choices between candidates supporting or opposing policies to extend transgender rights.

In addition, our findings on the mismatch between beliefs and actual views suggest that there may be scope for informational interventions to improve labor market outcomes for transgender individuals. Specifically, given that most respondents underestimate the overall level of support among the US population for transgender managers and employment non-discrimination laws protecting transgender individuals, informing individuals about the actual level of support for transgender individuals in the workplace could potentially shift individual's views, in line with other studies on gender norms (Bursztyn, González, and Yanagizawa-Drott 2020). If these mismatches between beliefs and actual views are not corrected, such misperceptions could lend legitimacy to anti-transgender policies that most people may not support.

Finally, our results indicate that transgender-specific labor market interventions may be necessary to achieve workplace equality for gender minorities, since individuals report significantly more positive views regarding LGB-related workplace support than transgender-related workplace support.

**References**

Agüero, Jorge M., and Veronica Frisancho. 2022. "Measuring Violence against Women with Experimental Methods." *Economic Development and Cultural Change* 70 (4): 1565–90.
Aksoy, Billur, Ian Chadd, and Boon Han Koh. 2021. "(Anticipated) Discrimination against Sexual

Minorities in Prosocial Domains." *Working Paper*.

Aksoy, Cevat G., Christopher S. Carpenter, Jeff Frank, and Matt L. Huffman. 2019. "Gay Glass Ceilings : Sexual Orientation and Workplace Authority in the UK." *Journal of Economic Behavior & Organization* 159 (March): 167–80.

Albrecht, James, Anders Björklund, and Susan Vroman. 2003. "Is There a Glass Ceiling in Sweden?" *Journal of Labor Economics* 21 (1): 145–77.

Altonji, Joseph G., and Rebecca M. Blank. 1999. "Race and Gender in the Labor Market." In *Handbook of Labor Economics*, edited by O. Ashenfelter and David Card, Volume 3, 3143–3259. Amsterdam: Elsevier Science.

Aronow, Peter m., Alexander Coppock, Forrest W. Crawford, and Donald P. Green. 2015. "Combining List Experiment and Direct Question Estimates of Sensitive Behavior Prevalence." *Journal of Survey Statistics and Methodology* 3 (1): 43–66.

Arrow, Kenneth. 1973. "The Theory of Discrimination." In *Discrimination in Labor Markets*. Princeton: Princeton University Press.

Badgett, M.V. Lee, Christopher S. Carpenter, and Dario Sansone. 2021. "LGBTQ Economics." *Journal of Economic Perspectives* 35 (2): 141–70.

Becker, Gary S. 1971. *The Economics of Discrimination*. 2nd ed. Chicago, IL: University of Chicago Press.

Berinsky, Adam J. 2004. *Silent Voices: Opinion Pools and Political Participation in America*. Princeton, NJ: Princeton University Press.

Bertrand, Marianne, and Esther Duflo. 2017. "Field Experiments on Discrimination." *Handbook of Economic Field Experiments* 1 (January): 309–93.

Bertrand, Marianne, and Sendhil Mullainathan. 2004. "Are Emily and Greg More Employable than Lakisha and Jamal? A Field Experiment on Labor Market Discrimination." *American Economic Review* 94 (4): 991–1013.

Bicchieri, Cristina. 2002. *The Grammar of Society: The Nature and Dynamics of Social Norms*. Cambridge University Press.

Blair, Graeme, and Kosuke Imai. 2012. "Statistical Analysis of List Experiments." *Political Analysis* 20 (1): 47–77.

Broockman, David, and Joshua Kalla. 2016. "Durably Reducing Transphobia: A Field Experiment on Door-to-Door Canvassing." *Science Magazine* 352 (6282): 220–24.

Brown, Anna. 2022a. "About 5% of Young Adults in the U.S. Say Their Gender Is Different from Their Sex Assigned at Birth." *Pew Research Center*.

———. 2022b. "Deep Partisan Divide on Whether Greater Acceptance of Transgender People Is Good for Society." *Pew Research Center*.

Bursztyn, Leonardo, Alessandra L. González, and David Yanagizawa-Drott. 2020. "Misperceived Social Norms: Women Working Outside the Home in Saudi Arabia." *American Economic Review* 110 (10): 2997–3029.

Carpenter, Christopher S., Samuel T. Eppink, and Gilbert Gonzales. 2020. "Transgender Status, Gender Identity, and Socioeconomic Outcomes in the United States." *ILR Review* 73 (3): 573–99.

Carpenter, Christopher S., Maxine J. Lee, and Laura Nettuno. 2022. "Economic Outcomes for Transgender People and Other Gender Minorities in the United States: First Estimates from a Nationally Representative Sample." *Southern Economic Journal* Accepted.

Chen, Daniel L., Martin Schonger, and Chris Wickens. 2016. "OTree-An Open-Source Platform for Laboratory, Online, and Field Experiments." *Journal of Behavioral and Experimental*

*Finance* 9: 88–97.

Chuang, Erica, Pascaline Dupas, Elise Huillery, and Juliette Seban. 2021. "Sex, Lies, and Measurement: Consistency Tests for Indirect Response Survey Methods." *Journal of Development Economics* 148 (January): 102582.

Coffman, Katherine B., Lucas C. Coffman, and Keith M. Marzilli Ericson. 2017. "The Size of the LGBT Population and the Magnitude of Antigay Sentiment Are Substantially Underestimated." *Management Science* 63 (10): 3168–86.

Coutts, Elisabeth, and Ben Jann. 2011. "Sensitive Questions in Online Surveys: Experimental Results for the Randomized Response Technique (RRT) and the Unmatched Count Technique (UCT)." *Sociological Methods & Research* 40 (1): 169–93.

Cullen, Zoë B., and Ricardo Perez-Truglia. 2021. "The Old Boys' Club: Schmoozing and the Gender Gap." *NBER Working Paper* 26530.

Delhommer, Scott. 2020. "Effect of State and Local Sexual Orientation Laws Anti-Discrimination Laws on Labor Market Differentials." *Working Paper*.

Detkova, Polina, Andrey Tkachenko, and Andrei Yakovlev. 2021. "Gender Heterogeneity of Bureaucrats in Attitude to Corruption: Evidence from List Experiment." *Journal of Economic Behavior & Organization* 189 (September): 217–33.

Doan, Long, Natasha Quadlin, and Brian Powell. 2022. "Attitudes Toward Formal Rights and Informal Privileges for Transgender People: Evidence from a National Survey Experiment." In *Demography of Transgender, Nonbinary and Gender Minority Populations*, edited by Amanda K. Baumle and Sonny Nordmarken, 47–72. Cham ,Switzerland: Springer, Cham.

Donohue, John J. III, and James J. Heckman. 1991. "Continuous Versus Episodic Change: The Impact of Civil Rights Policy on the Economic Status of Black." *Journal of Economic Literature* 29 (4): 1603–43.

Droitcour, Judith, Rachel A. Caspar, Michael L. Hubbard, Teresa L. Parsley, Wendy Visscher, and Trena M. Ezzati. 1991. "The Item Count Technique as a Method of Indirect Questioning: A Review of Its Development and a Case Study Application." In *Measurement Errors in Surveys*, edited by P. P. Biemer, R. M. Groves, L. E. Lyberg, N. A. Mathiowetz, and S. Sudman, 185–210. New York, NY: John Wiley & Sons, Ltd.

Eyal, Peer, Rothschild David, Gordon Andrew, Evernden Zak, and Damer Ekaterina. 2021. "Data Quality of Platforms and Panels for Online Behavioral Research." *Behavior Research Methods* 54: 1643–1662.

Flores, Andrew R., Patrick Miller, and Barry Tadlock. 2018. "Public Opinion about Transgender People and Policies." In *The Remarkable Rise of Transgender Rights*, edited by Jami K. Taylor, Daniel C. Lewis, and Donald P. Haider-Markel, 74. University of Michigan Press.

Fouka, Vasiliki, and M. A.R.C.O. Tabellini. 2022. "Changing In-Group Boundaries: The Effect of Immigration on Race Relations in the United States." *American Political Science Review* 116 (3): 968–84.

Frank, Jeff. 2006. "Gay Glass Ceilings." *Economica* 73 (291): 485–508.

Gallup. 2022. "In Depth: LGBT Rights."

Geijtenbeek, Lydia, and Erik Plug. 2018. "Is There a Penalty for Registered Women? Is There a Premium for Registered Men? Evidence from a Sample of Transsexual Workers." *European Economic Review* 109 (October): 334–47.

Gervais, Will M., and Maxine B. Najle. 2018. "How Many Atheists Are There?" *Social Psychological and Personality Science* 9 (1): 3–10.

Giuliano, Laura, David I. Levine, and Jonathan Leonard. 2009. "Manager Race and the Race of New Hires." *Journal of Labor Economics* 27 (4): 589–631.

Glynn, Adam N. 2013. "What Can We Learn with Statistical Truth Serum? Design and Analysis of the List Experiment." *Public Opinion Quarterly* 77 (S1): 159–72.

Goldin, Claudia, and Cecilia Rouse. 2000. "Orchestrating Impartiality: The Impact of 'Blind' Auditions on Female Musicians." *The American Economic Review* 90 (4): 715–41.

Granberg, Mark, Per A. Andersson, and Ali Ahmed. 2020. "Hiring Discrimination Against Transgender People: Evidence from a Field Experiment." *Labour Economics* 65 (August): 101860.

GSS. 2021. "Political Party Affiliation." *GSS Data Explorer*.

Gupta, Neeraja, Luca Rigotti, and Alistair Wilson. 2021. "The Experimenters' Dilemma: Inferential Preferences over Populations." *Working Paper*.

Holbrook, Allyson L., and Jon A. Krosnick. 2010. "Social Desirability Bias in Voter Turnout Reports: Tests Using the Item Count Technique." *Public Opinion Quarterly* 74 (1): 37–67.

Isler, Ozan, John Maule, and Chris Starmer. 2018. "Is Intuition Really Cooperative? Improved Tests Support the Social Heuristics Hypothesis." *PLOS ONE* 13 (1): e0190560.

Jamison, Julian, Dean S. Karlan, and Pia Raffler. 2013. "Mixed Method Evaluation of a Passive Health Sexual Information Texting Service in Uganda." *Information Technologies & International Development* 9 (3): 1–28.

John, Leslie K., George Loewenstein, Alessandro Acquisti, and Joachim Vosgerau. 2018. "When and Why Randomized Response Techniques (Fail to) Elicit the Truth." *Organizational Behavior and Human Decision Processes* 148 (September): 101–23.

Jones, Jeffrey M. 2022. "LGBT Identification in U.S. Ticks Up to 7.1%." *Gallup*.

Kane, James G., Stephen C. Craig, and Kenneth D. Wald. 2004. "Religion and Presidential Politics in Florida: A List Experiment." *Social Science Quarterly* 85 (2): 281–93.

Karlan, Dean S., and Jonathan Zinman. 2012. "List Randomization for Sensitive Behavior: An Application for Measuring Use of Loan Proceeds." *Journal of Development Economics* 98 (1): 71–75.

Klawitter, Marieka. 2011. "Multilevel Analysis of the Effects of Antidiscrimination Policies on Earnings by Sexual Orientation." *Journal of Policy Analysis and Management* 30 (2): 334–58.

———. 2015. "Meta-Analysis of the Effects of Sexual Orientation on Earnings." *Industrial Relations: A Journal of Economy and Society* 54 (1): 4–32.

Klawitter, Marieka, and Victor Flatt. 1998. "The Effects of State and Local Antidiscrimination Policies on Earnings for Gays and Lesbians." *Journal of Policy Analysis and Management* 17 (4): 658–86.

Kuklinski, James H., Michael D. Cobb, and Martin Gilens. 1997. "Racial Attitudes and the 'New South.'" *Journal of Politics* 59 (2): 323–49.

Kuklinski, James H., Paul M. Sniderman, Kathleen Knight, Thomas Piazza, Philip E. Tetlock, Gordon R. Lawrence, and Barbara Mellers. 1997. "Racial Prejudice and Attitudes Toward Affirmative Action." *American Journal of Political Science* 41 (2): 402.

LaBrie, Joseph W., and Mitchell Earleywine. 2000. "Sexual Risk Behaviors and Alcohol: Higher Base Rates Revealed Using the Unmatched-count Technique." *The Journal of Sex Research* 37 (4): 321–26.

Lang, Kevin, and Ariella Kahn-Lang Spitzer. 2020. "Race Discrimination: An Economic Perspective." *Journal of Economic Perspectives* 34 (2): 68–89.

Lax, Jeffrey R., Justin H. Phillips, and Alissa F. Stollwerk. 2016. "Are Survey Respondents Lying

about Their Support for Same-Sex Marriage? Lessons from a List Experiment." *Public Opinion Quarterly* 80 (2): 510–33.

Lewis, Daniel C., Andrew R. Flores, Donald P. Haider-Markel, Patrick R. Miller, Barry L. Tadlock, and Jami K. Taylor. 2017. "Degrees of Acceptance: Variation in Public Attitudes toward Segments of the LGBT Community:" *Political Research Quarterly* 70 (4): 861–75.

Lewis, Daniel C, Andrew R Flores, Donald P Haider-Markel, Patrick R Miller, and Jami K Taylor. 2022. "Transitioning Opinion?Assessing the Dynamics of Public Attitudes Toward Transgender Rights." *Public Opinion Quarterly*, May.

Luhur, Winston, Taylor N. T. Brown, and Andrew R. Flores. 2019. "Public Opinion of Transgender Rights in the United States." *Williams Institute*, 1–28.

Matsa, David A., and Amalia R. Miller. 2011. "Chipping Away at the Glass Ceiling: Gender Spillovers in Corporate Leadership." *American Economic Review* 101 (3): 635–39.

McCarthy, Justin. 2021. "Mixed Views Among Americans on Transgender Issues." *Gallup*, 1–6.

McKenzie, David, and Melissa Siegel. 2013. "Eliciting Illegal Migration Rates through List Randomization." *Migration Studies* 1 (3): 276–91.

Miller, Judith Droitcour. 1984. "A New Survey Technique for Studying Deviant Behavior." *Dissertation*, 1–198.

Morisi, Davide, John T. Jost, and Vishal Singh. 2019. "An Asymmetrical 'President-in-Power' Effect." *American Political Science Review* 113 (2): 614–20.

National Academies of Sciences Engineering and Medicine. 2020. "Understanding the Well-Being of LGBTQI+ Populations." Washington, D.C.

Neumark, David. 2018. "Experimental Research on Labor Market Discrimination." *Journal of Economic Literature* 56 (3): 799–866.

Neumark, David, and Wendy A. Stock. 2006. "The Labor Market Effects of Sex and Race Discrimination Laws." *Economic Inquiry* 44 (3): 385–419.

Oreffice, Sonia, and Climent Quintana-Domeque. 2021. "Gender Inequality in COVID-19 Times: Evidence from UK Prolific Participants." *Journal of Demographic Economics* 87 (2): 261–87.

Osman, Adam, Jamin D. Speer, and Andrew Weaver. 2021. "Discrimination Against Women in Hiring." *Working Paper*.

Palan, Stefan, and Christian Schitter. 2018. "Prolific.Ac—A Subject Pool for Online Experiments." *Journal of Behavioral and Experimental Finance* 17 (March): 22–27.

Peer, Eyal, Laura Brandimarte, Sonam Samat, and Alessandro Acquisti. 2017. "Beyond the Turk: Alternative Platforms for Crowdsourcing Behavioral Research." *Journal of Experimental Social Psychology* 70 (May): 153–63.

Phelps, Edmund S. 1972. "The Statistical Theory of Racism and Sexism." *American Economic Review* 62 (4): 659–61.

Robertson, Ronald E., Felix W. Tran, Lauren N. Lewark, and Robert Epstein. 2018. "Estimates of Non-Heterosexual Prevalence: The Roles of Anonymity and Privacy in Survey Methodology." *Archives of Sexual Behavior* 47 (4): 1069–84.

Ruggles, Steven, Sarah Flood, Ronald Goeken, Megan Schouweiler, and Matthew Sobek. 2022. "IPUMS USA: Version 12.0 [Dataset]." *IPUMS*.

Schild, Christoph, Daniel W. Heck, Karolina A. Ścigała, and Ingo Zettler. 2019. "Revisiting REVISE: (Re)Testing Unique and Combined Effects of REminding, VIsibility, and SElf-Engagement Manipulations on Cheating Behavior." *Journal of Economic Psychology* 75 (A): 102161.

Shannon, Matthew. 2022. "The Labour Market Outcomes of Transgender Individuals." *Labour Economics* 77: 102006.

Streb, Matthew J., Barbara Burrell, Brian Frederick, and Michael A. Genovese. 2008. "Social Desirability Effects and Support for a Female American President." *Public Opinion Quarterly* 72 (1): 76–89.

Taylor, Jami K., Daniel C. Lewis, and Donald P. Haider-Markel. 2018. *The Remarkable Rise of Transgender Rights*. University of Michigan Press.

Tsai, Chi Lin. 2019. "Statistical Analysis of the Item-Count Technique Using Stata." *Stata Journal* 19 (2): 390–434.

Tsuchiya, Takahiro, Yoko Hirai, and Shigeru Ono. 2007. "A Study of the Properties of the Item Count Technique." *Public Opinion Quarterly* 71 (2): 253–72.

U.S. Census. 2021. "Quick Facts." Washington, D.C. https://www.census.gov/quickfacts/fact/table/US/PST045221.

Zmigrod, Leor, Peter J. Rentfrow, and Trevor W. Robbins. 2018. "Cognitive Underpinnings of Nationalistic Ideology in the Context of Brexit." *Proceedings of the National Academy of Sciences* 115 (19): E4532–40.

**Figure 1: List experiments.**

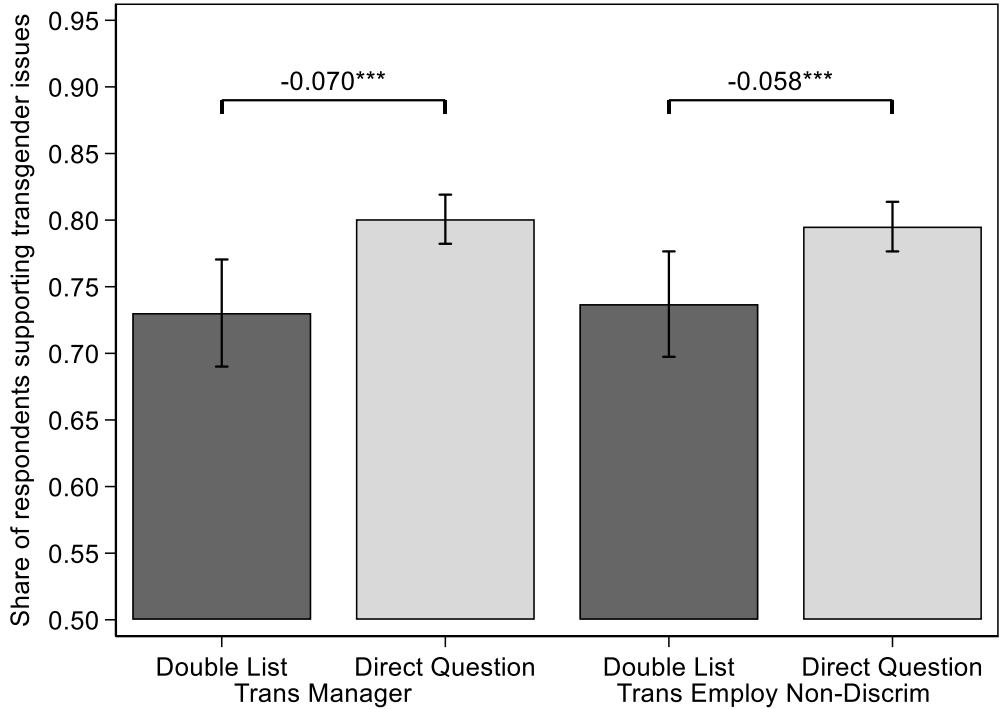

* p < 0.10, ** p < 0.05, *** p < 0.01. 95-percent confidence intervals reported with vertical range plots. The numbers above the horizontal bars are the differences between the two groups at the base of each horizontal bar. Trans Manager key statement: "I would be comfortable having a transgender manager at work". Trans Employ Non-Discrim key statement: "I think the law should prohibit employment discrimination against transgender individuals". Number of observations: 1,806. Source: 2022 Prolific List Experiment. See also Figure B3 and Table B3.

**Figure 2: Perceptions of general views.**

**Panel A: Respondent thinks X/100 would be comfortable having a transgender manager at work.**

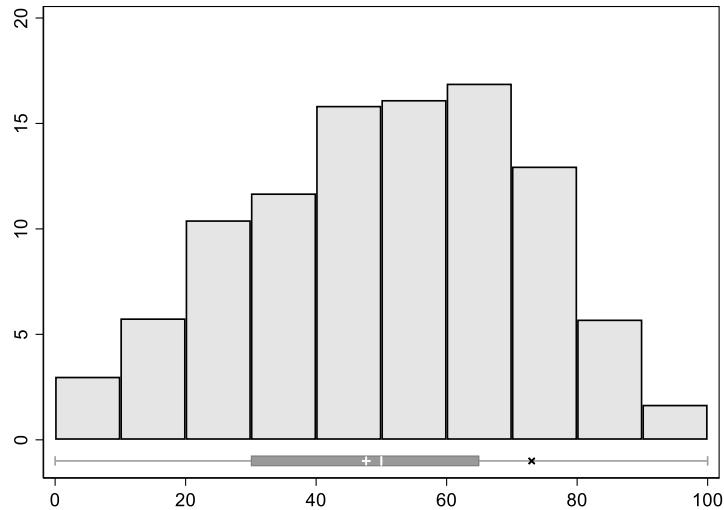

**Panel B: Respondent thinks X/100 would agree that the law should prohibit employment discrimination against transgender individuals.**

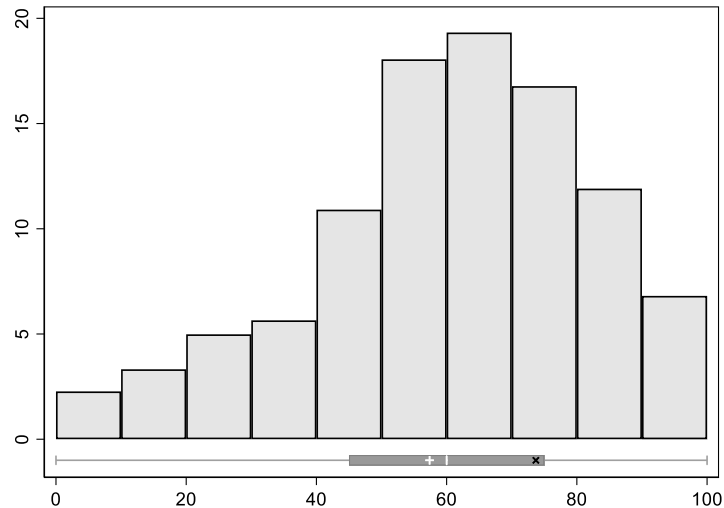

The original survey question for Panel A is "Out of every 100 people in the general US population, I think approximately __ out of 100 would be comfortable with having a transgender manager at work." The original survey question for Panel B is "Out of every 100 people in the general US population, I think approximately __ out of 100 would agree that the law should prohibit employment discrimination against transgender individuals." The box plot below each histogram reports minimum and maximum values, 25th and 75th percentiles, as well as mean and median. Within each box plot, the white vertical line " | " indicates the median, the white " + " symbol indicates the mean. The black " X " symbol in Panel A indicates the actual share of the sample being comfortable with a transgender manager estimated from the double-list experiment, while in Panel B indicates the actual share of the sample agreeing that the law should prohibit employment discrimination against transgender individuals estimated from the double-list experiment (see Figure 1). Number of observations: 1,806. Source: 2022 Prolific List Experiment.

**Figure 3: List experiment on transgender manager. Difference-in-means comparisons. Heterogeneity analysis.**

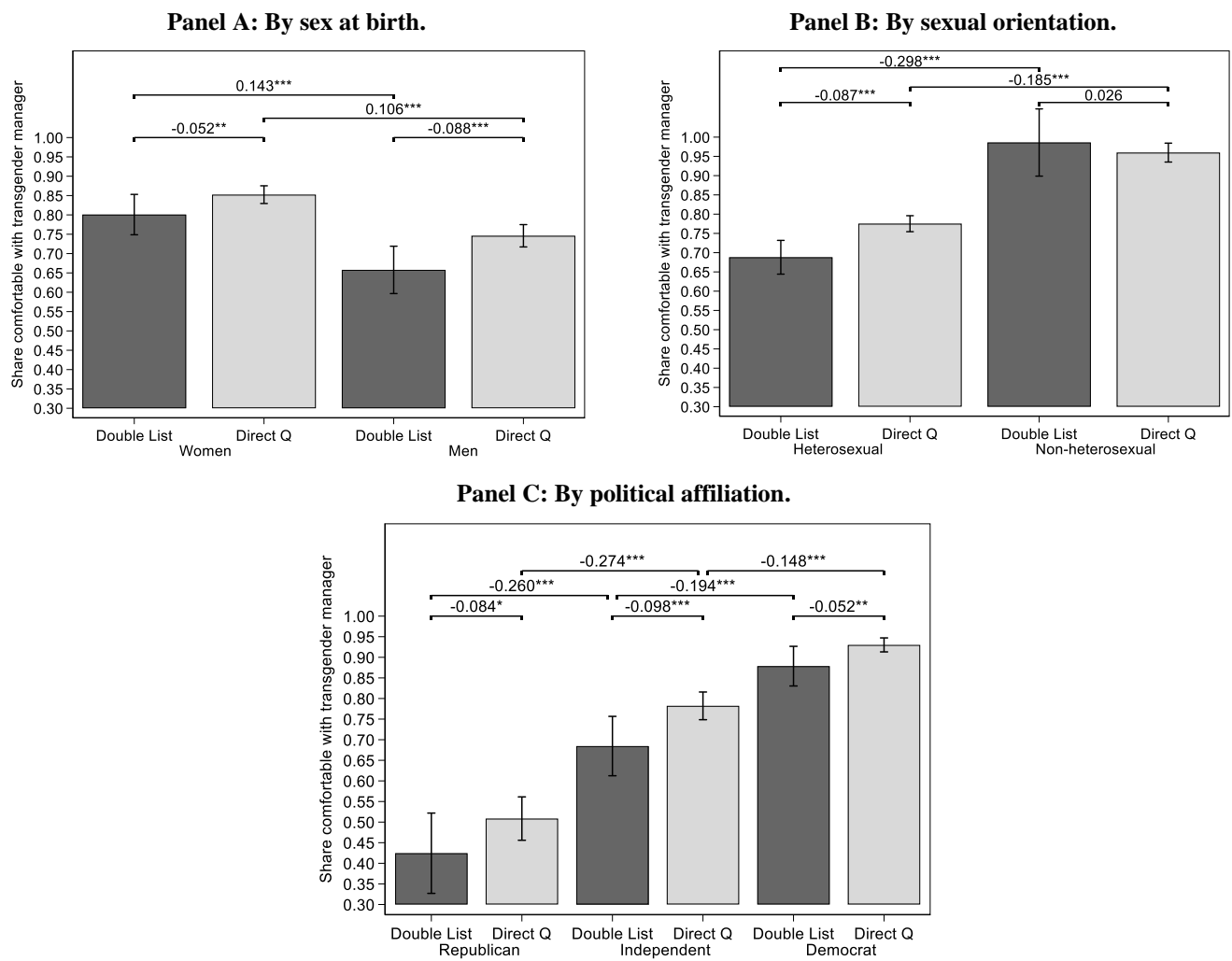

* p < 0.10, ** p < 0.05, *** p < 0.01. 95-percent confidence intervals reported with vertical range plots. The numbers above the horizontal bars in each figure are the differences between the two groups at the base of each horizontal bar. Key statement: "I would be comfortable having a transgender manager at work". Number of observations: 1,806. Source: 2022 Prolific List Experiment. See also Tables B4-B6.

**Figure 4: List experiment on transgender employment non-discrimination protection. Difference-in-means comparisons. Heterogeneity analysis.**

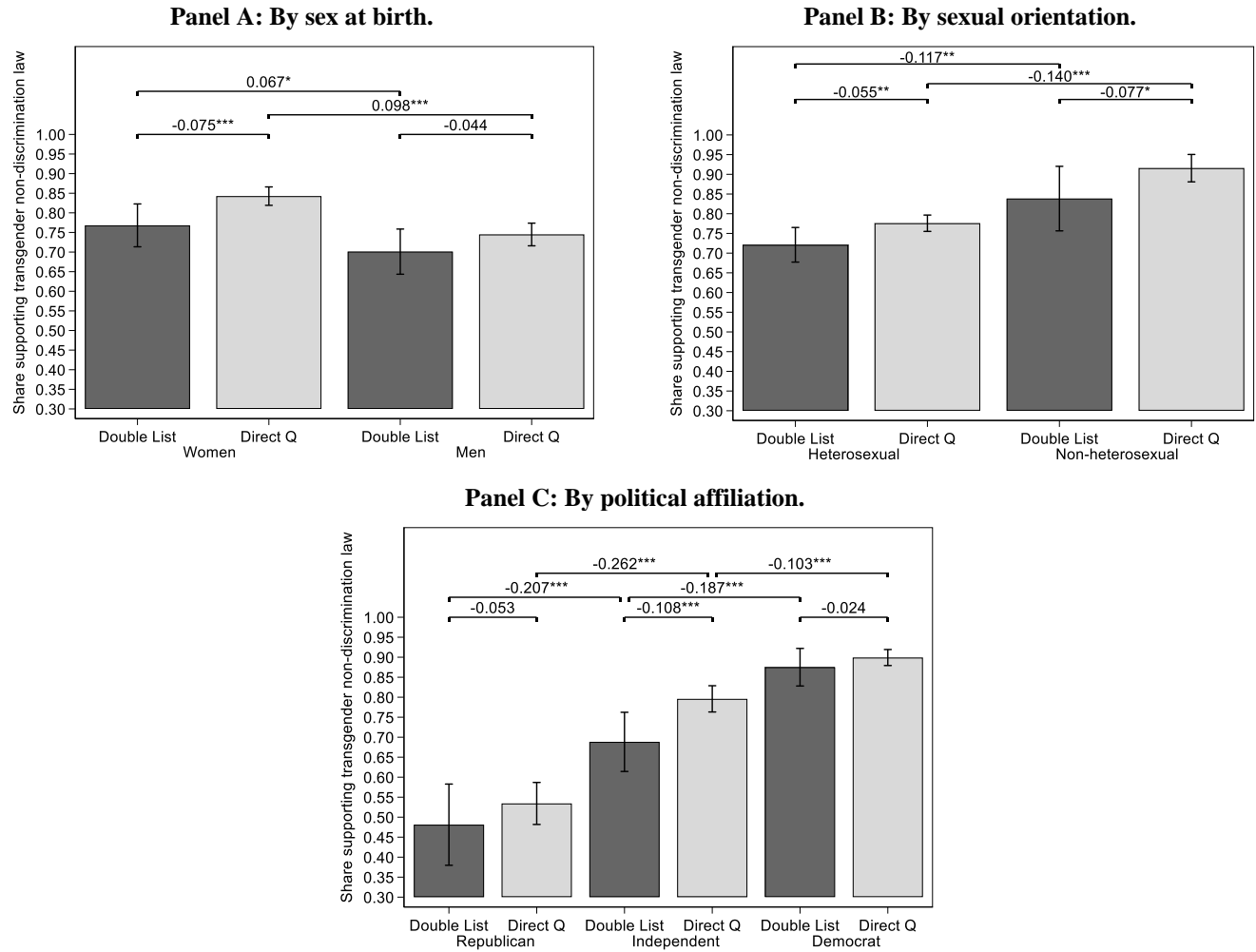

* $p < 0.10$, ** $p < 0.05$, *** $p < 0.01$. 95-percent confidence intervals reported with vertical range plots. The numbers above the horizontal bars in each figure are the differences between the two groups at the base of each horizontal bar. Key statement: "I think the law should prohibit employment discrimination against transgender individuals". Number of observations: 1,806. Source: 2022 Prolific List Experiment. See also Tables B4-B6.

**Figure 5: Comparison of views toward transgender individuals relative to LGB individuals and issues.**

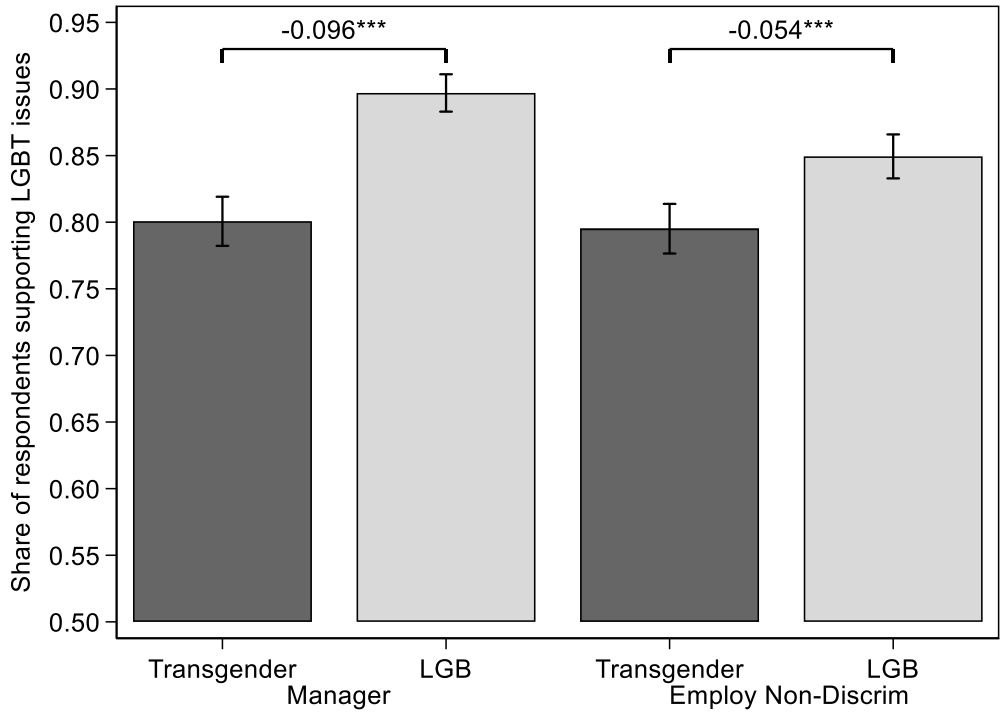

* p < 0.10, ** p < 0.05, *** p < 0.01. 95-percent confidence intervals reported with horizontal range plots. The numbers above the horizontal bars are the differences between the two groups at the base of each horizontal bar. Questions used in this table are the following for "Manager": "Would you be comfortable having a [*transgender*] / [*openly lesbian, gay, or bisexual*] manager at work?". For "Employ Non-Discrim": "Do you think the law should prohibit employment discrimination against [*transgender*] / [*openly lesbian, gay, or bisexual*] individuals?". Number of observations: 1,806. Source: 2022 Prolific List Experiment.

**Table 1: List experiment example.**

| Short List | Long List |
|---|---|
| • I have a driver's license<br>• I think COVID-19 health risks were overstated<br>• I can fluently speak at least three languages<br>• I support the Black Lives Matter movement | • I have a driver's license<br>• I think COVID-19 health risks were overstated<br>• I can fluently speak at least three languages<br>• I support the Black Lives Matter movement<br>• I would be comfortable having a transgender manager at work *[key statement]* |

The order of the statements within each list was randomized at the subject level. For the full set of lists, see Online Appendix C.

**Table 2: Summary statistics of participant characteristics.**

| Variable | Mean |
|---|---|
| Age | |
| *Mean* | 44.74 |
| *between 18-34* | 0.334 |
| *between 35-49* | 0.254 |
| *between 50-64* | 0.282 |
| *65 or older* | 0.130 |
| Female (sex at birth) | 0.514 |
| Race | |
| *White only* | 0.745 |
| *Black or African American only* | 0.135 |
| *Asian or Native Hawaiian or Pacific Islander only* | 0.065 |
| Married | 0.441 |
| Education | |
| *High School, GED, or less* | 0.107 |
| *Some College credits, no degree* | 0.200 |
| *Associate's degree* | 0.110 |
| *Bachelor's degree or higher* | 0.583 |
| Employed | 0.670 |
| Household income: less than $60,000 | 0.477 |
| Political Party Affiliation | |
| *Democrat* | 0.483 |
| *Republican* | 0.194 |
| *Independent* | 0.323 |
| Urbanicity | |
| *Rural area* | 0.126 |
| *Small city or town* | 0.291 |
| *Suburb near a large city* | 0.348 |
| *Large city* | 0.236 |
| Region | |
| *Northeast* | 0.211 |
| *Midwest* | 0.215 |
| *South* | 0.424 |
| *West* | 0.150 |
| Total number of participants | 1,806 |

Race categories are not mutually exclusive (participants could select more than one option). The variable "Employed" includes both "employed for wages" and "self-employed". Source: 2022 Prolific List Experiment.

**Table 3: List experiments. Multivariate analysis.**

|  | List A | | | | List B | | | |
|---|---|---|---|---|---|---|---|---|
|  | (1) | (2) | (3) | (4) | (5) | (6) | (7) | (8) |
| *Panel A: Transgender manager* | | | | | | | | |
| Subject saw list with key statement | 0.719*** | 0.718*** | 0.718*** | 0.717*** | 0.741*** | 0.728*** | 0.724*** | 0.721*** |
|  | (0.030) | (0.030) | (0.029) | (0.029) | (0.034) | (0.033) | (0.033) | (0.033) |
| $R^2$ | 0.239 | 0.280 | 0.351 | 0.356 | 0.212 | 0.263 | 0.294 | 0.304 |
| | | | | | | | | |
| *Panel B: Trans employment non-discrimination protection* | | | | | | | | |
| Subject saw list with key statement | 0.734*** | 0.724*** | 0.727*** | 0.729*** | 0.740*** | 0.748*** | 0.749*** | 0.751*** |
|  | (0.030) | (0.031) | (0.030) | (0.030) | (0.031) | (0.031) | (0.030) | (0.030) |
| $R^2$ | 0.247 | 0.277 | 0.296 | 0.297 | 0.243 | 0.294 | 0.332 | 0.338 |
| | | | | | | | | |
| *Controls for:* | | | | | | | | |
| State FE | | ✓ | ✓ | ✓ | | ✓ | ✓ | ✓ |
| Demographic controls | | ✓ | ✓ | ✓ | | ✓ | ✓ | ✓ |
| Socio-economic factors and beliefs | | | ✓ | ✓ | | | ✓ | ✓ |
| Additional controls | | | | ✓ | | | | ✓ |
| Observations | 1,806 | 1,806 | 1,806 | 1,806 | 1,806 | 1,806 | 1,806 | 1,806 |

* $p < 0.10$, ** $p < 0.05$, *** $p < 0.01$. Robust standard errors clustered in parentheses. Transgender manager key statement: "I would be comfortable having a transgender manager at work". Trans employment non-discrimination protection key statement: "I think the law should prohibit employment discrimination against transgender individuals". *Demographic controls* include subject's age, sex at birth, race (including missing indicator), sexual orientation, and sexual attraction. *Socio-economic factors and beliefs* include subject's education level, employment status, income, current religious affiliation, political affiliation, and beliefs about general level of support for transgender managers (Panel A) or employment discrimination protection for transgender individuals (Panel B). *Additional controls* include whether at least one child less than 18 years of age lives in the subject's household, number of people living in the subject's household, urbanicity, and marital status. OLS estimates. Source: 2022 Prolific List Experiment.

**Table 4: List experiments. Heterogeneity analysis. Multivariate analysis.**

|  | Transgender manager | Trans employment non-discrimination protection |
|---|---|---|
|  | (1) | (2) |
| *Interaction of treatment variable with:* |  |  |
| Sex assigned at birth: Female | 0.093** | 0.052 |
|  | (0.040) | (0.040) |
| Race: White only | 0.020 | -0.031 |
|  | (0.048) | (0.047) |
| Age: 18-44 | 0.065 | 0.048 |
|  | (0.042) | (0.042) |
| Sexual orientation: Heterosexual | -0.233*** | 0.029 |
|  | (0.068) | (0.065) |
| Sexual attraction: Different-sex only | -0.004 | -0.052 |
|  | (0.059) | (0.060) |
| Political affiliation: Republican | -0.326*** | -0.342*** |
|  | (0.060) | (0.062) |
| Political affiliation: Independent or Other | -0.161*** | -0.179*** |
|  | (0.043) | (0.045) |
| Household income: Less than $60,000 | -0.021 | 0.012 |
|  | (0.040) | (0.040) |
| Education: Less than a Bachelor's degree | -0.089** | 0.033 |
|  | (0.041) | (0.043) |
| Employment status: Employed for wages | -0.063 | -0.022 |
|  | (0.042) | (0.044) |
| Current religious affiliation: Christian | -0.024 | -0.022 |
|  | (0.078) | (0.074) |
| Current religious affiliation: Not religious | 0.053 | 0.045 |
|  | (0.076) | (0.072) |
| Currently live in: North-East | 0.029 | 0.008 |
|  | (0.051) | (0.051) |
| Currently live in: Midwest | -0.008 | 0.007 |
|  | (0.050) | (0.052) |
| Currently live in: West | -0.053 | 0.095 |
|  | (0.059) | (0.059) |
| Respondent believes 50% or more of Americans would be comfortable with a transgender manager at work | 0.182*** |  |
|  | (0.039) |  |
| Respondent believes 50% or more of Americans would agree that the law should prohibit employment discrimination against transgender individuals |  | 0.096** |
|  |  | (0.045) |
| Constant | 0.924*** | 0.752*** |
|  | (0.108) | (0.108) |
| Observations | 1,806 | 1,806 |

* $p < 0.10$, ** $p < 0.05$, *** $p < 0.01$. Robust standard errors clustered in parentheses. Transgender manager key statement: "I would be comfortable having a transgender manager at work". Trans employment non-discrimination protection key statement: "I think the law should prohibit employment discrimination against transgender individuals". Coefficients obtained using the Stata command *kict ls* (Tsai, 2019) performing least squares estimation for a double list experiment. The dependent variables are the reported true number of statements for the transgender manager lists (Column 1) and the employment non-discrimination protection lists (Column 2). The treatment variable is an indicator variable equal to 1 for the first long list (List A) containing the corresponding key statement and the second short list (List B), 0 for the first short list (List A) and the second long list (List B). All estimated coefficients of the interactions of the treatment variable with the observable characteristics are reported except for the variable "missing race".

# Online Appendix (NOT MEANT FOR PUBLICATION)

## Appendix A. Experimental design details and quality checks

### A1. Experimental design details

Although it is common practice in the literature not to randomize the order of the lists, we chose to incorporate some randomization into our design to control for potential order effects (here, we refer to the order of the lists, not the order of the statements within the list). More specifically, we created the following four paths that a participant follows:

**Path 1** - (Manager List A), (Manager List B + KS 1), (Employ Non-Discrim List A), (Employ Non-Discrim List B + KS 2)

**Path 2** - (Manager List A + KS 1), (Manager List B), (Employ Non-Discrim List A + KS 2), (Employ Non-Discrim List B)

**Path 3** - (Employ Non-Discrim List B), (Employ Non-Discrim List A + KS 2), (Manager List B), (Manager List A + KS 1)

**Path 4** - (Employ Non-Discrim List B + KS 2), (Employ Non-Discrim List A), (Manager List B + KS 1), (Manager List A)

KS 1 and KS 2 stand for transgender manager key statement and transgender employment non-discrimination protection key statement, respectively. Manager List A, Manager List B, Employ Non-Discrim List A, and Employ Non-Discrim List B can be seen in the instructions in Online Appendix C. As can be seen above, half of our participants saw List As first, and the other half saw List Bs first. When we compare the distribution of answers across these two orders using Pearson's chi-square test (i.e., comparing responses in Path 1 to Path 4 and Path 2 to Path 3), we do not see any significant differences between the lists.

### A2. Further quality checks

#### A.2.1. Data quality checks

As discussed in Section 3.2., we carefully constructed each list to avoid floor and ceiling effects (i.e., participants reporting zero items or all items, thus removing the privacy protection provided by the list experiment). We check for ceiling and floor effects and present findings in Figures B1-B2. As can be seen in these figures, only a very small share of our participants reports the highest and lowest possible items in each of the lists. Thus, we conclude that the floor and ceiling effects are negligible in our experiment. Additionally, if the distributions of responses had followed a uniform distribution, then it would have indicated that most respondents provided random answers (Coffman, Coffman, and Ericson 2017). As shown in Figures B1 and B2, it is therefore reassuring to note that our distributions of responses do not follow such a uniform distribution.

Next, we check the robustness of our main list experiment findings by excluding participants who completed the study very quickly or very slowly since they may not be paying as much attention to the study instructions. On average, it took 420 seconds (7 minutes) to complete the experiment. We exclude a total of 183 participants who took less than 211 seconds (top 5%) and those who took more than 796 seconds (bottom 5%). The results are presented in Table B3 Panel D and show that our findings are robust to removing these participants.

Following our pre-analysis plan, we also checked if some respondents provided the same number for all list experiments (which might be an indication of participants not paying attention). Across all five lists, nobody provided the same number. Looking at the first four lists (thus excluding the list that serves as an attention check), 64 participants provided the same number for all four lists. Our main findings (Figure 1 and Table B3) are robust to the exclusion of these 64 participants.

### A.2.2. List experiment assumptions

The validity of a list experiment relies on three assumptions: 1) treatment randomization, 2) no design effect, and 3) no liar. The first assumption means that the sample is split at random. The second assumption means that respondents do not give different answers to non-key statements depending on whether they are in the long list group. The third assumption means that respondents answer the key statement truthfully.

A common practice to check the first assumption - treatment randomization - is to test for differences between the short list and long list groups' responses to important variables in the survey. We do this in Table B18 where we check the differences between the two groups in terms of their demographic covariates. We do not see a significant difference between the two groups except for sex where one group has slightly more females than the other. We conclude that our randomization of treatment was effective. Moreover, following Gerber and Green (2012) and Detkova, Tkachenko, and Yakovlev (2021), we do not only rely on means comparisons but also employ regression analyses where we control for observable characteristics (as discussed in Section 4.1).

The second assumption – no design effect – requires respondents not to change their answers to non-key statements depending on whether the key statement appears in the list (i.e., whether they see the long list). To clarify, suppose that a respondent in the short list group answers two non-key statements affirmatively. If they were assigned to the long list group, their answer must be either '2' or '3' (that is, they either answer two non-key statements affirmatively or answer two non-key statements plus the key statement affirmatively). It is worth noting that we do not assume that subjects give truthful answers to these non-key statements, we only assume that the answers are consistent in short and long list groups. Blair and Imai (2012) proposed a statistical test for the no-design-effect assumption. The first step is to estimate the probabilities of all possible types of item-count responses. If some of these estimated probabilities were a nonsensical value (e.g., a negative value), it would raise doubts about the validity of the no-design-effect assumption. One can then test whether such negative estimates have arisen by chance. In our two list experiments regarding transgender managers (Lists 1A and 1B), none of the estimated probabilities is below zero or above one. The same can be said about List 2A regarding employment non-discrimination protection. For List 2B regarding employment non-discrimination protection, two out of the ten estimated probabilities are slightly below zero.[30] Nevertheless, one cannot reject the null that such estimates have arisen by chance. Therefore, it is possible to conclude that the available evidence supports the "no design effect" assumption.

It is not statistically feasible to check the 'no liar' assumption, not only because respondents' answers to the key statement are by design unobserved, but also because their truthful answers are unknown (otherwise there would be no point in using the list experiment technique). By running this experiment in an online anonymized platform and by making sure when designing the lists that agreeing to all or none of the statement is highly unlikely, we have tried to limit any concerns about this assumption. Indeed, Figures B1 and B2 present the distribution of responses for each list and key statements: the modal response in all lists is 2. Moreover, as noted in the previous section, the percentage of times where the responses are 0 or 4 (5 for long lists) is negligible, meaning that the privacy of responses was protected.

**References in Online Appendix A**

Blair, Graeme, and Kosuke Imai. 2012. "Statistical Analysis of List Experiments." *Political Analysis* 20 (1): 47–77.
Coffman, Katherine B., Lucas C. Coffman, and Keith M. Marzilli Ericson. 2017. "The Size of the LGBT Population and the Magnitude of Antigay Sentiment Are Substantially Underestimated." *Management Science* 63 (10): 3168–86.
Detkova, Polina, Andrey Tkachenko, and Andrei Yakovlev. 2021. "Gender Heterogeneity of Bureaucrats in Attitude to Corruption: Evidence from List Experiment." *Journal of Economic Behavior & Organization* 189 (September): 217–33.
Gerber, Alan S., and Donald P. Green. 2012. *Field Experiments: Design, Analysis, and Interpretation*. New York, NY: W. W. Norton & Company.
Tsai, Chi Lin. 2019. "Statistical Analysis of the Item-Count Technique Using Stata." *Stata Journal* 19 (2): 390–434.

---

[30] We use the Stata command *kict deff* (Tsai 2019). For Lists 2A and 2B, since no respondent answered "5" when provided with the long list, the command was not able to distinguish between the long list and the short list. Therefore, in order to conduct this test, we increased the number of items in Lists 2A and 2B reported by one respondent, randomly chosen, from "4" to "5". Our conclusions do not change when we randomly choose different respondents.

# Appendix B. Additional figures and tables

**Figure B1: Distribution of responses by list. Transgender manager.**

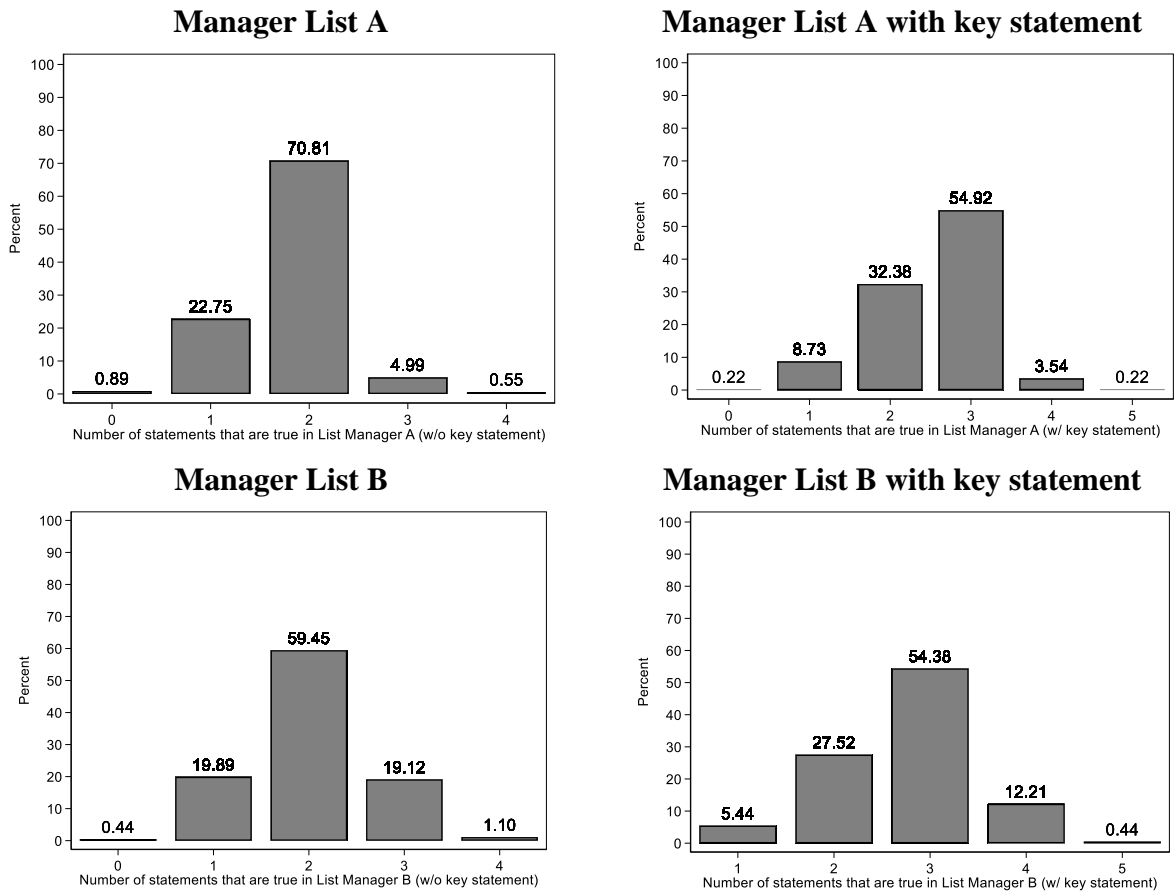

Key statement in the list: "I would be comfortable with having a transgender manager at work." Number of observations: 1,806. Source: 2022 Prolific List Experiment.

**Figure B2: Distribution of responses by list. Transgender employment non-discrimination protection.**

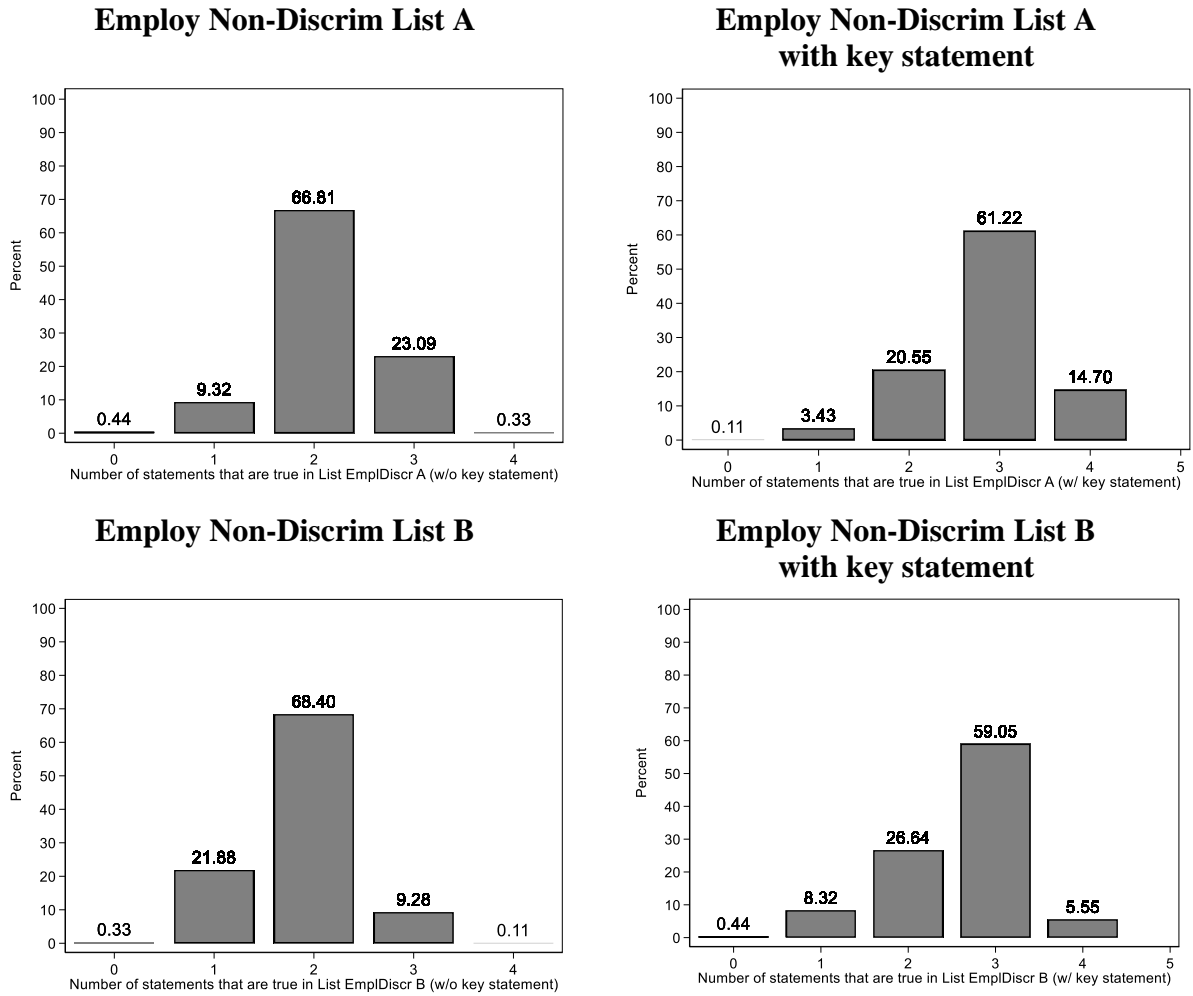

Key statement in the list: "I think the law should prohibit employment discrimination against transgender individuals." Number of observations: 1,806. Source: 2022 Prolific List Experiment.

**Figure B3: Main list experiments including List A and List B.**

**Panel A: Transgender managers.**

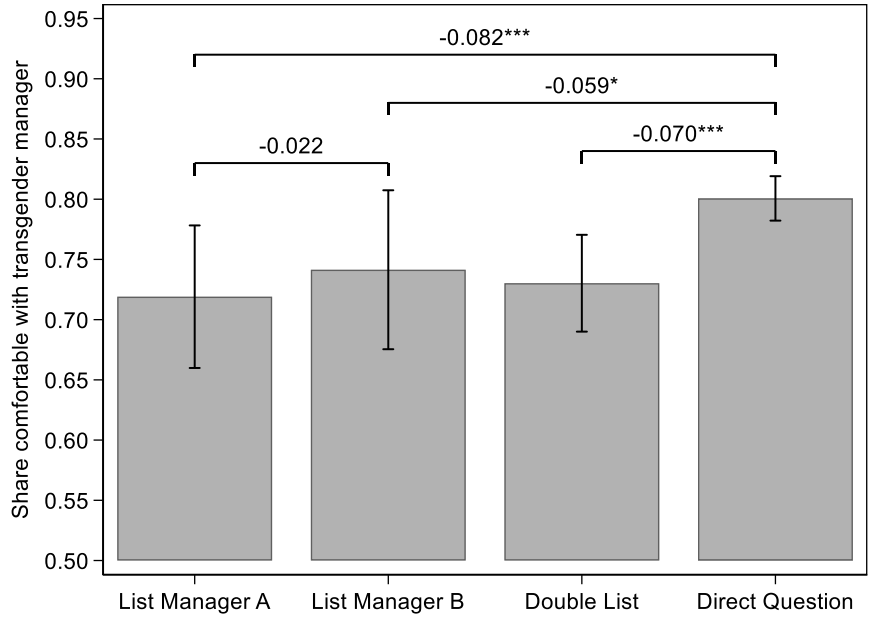

**Panel B: Employment non-discrimination protection for transgender individuals.**

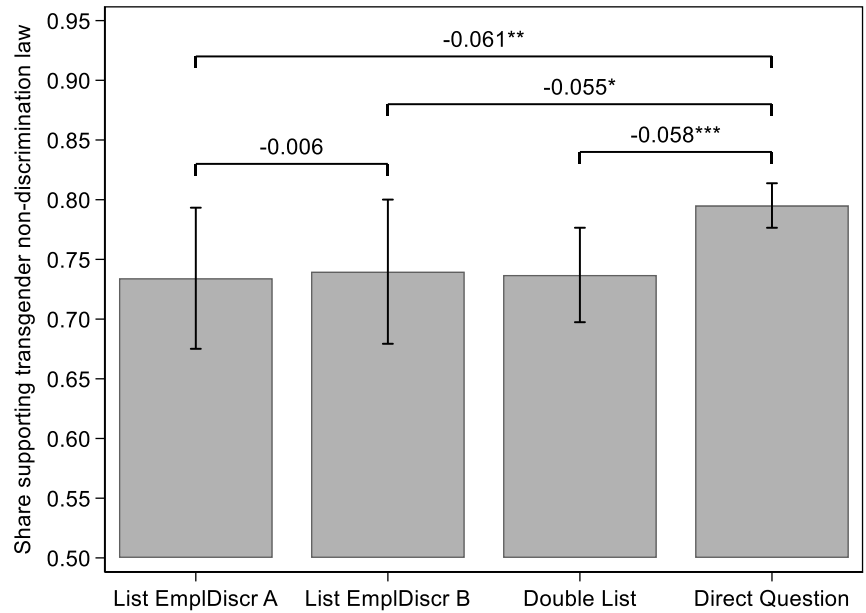

* p < 0.10, ** p < 0.05, *** p < 0.01. 95-percent confidence intervals reported with vertical range plots. The numbers above the horizontal bars are the differences between the two groups at the base of each horizontal bar. Number of observations: 1,806. Source: 2022 Prolific List Experiment.

**Figure B4: Respondent thinks X/100 would be comfortable having a transgender manager at work. By responses to direct question.**

**Panel A: Respondent would be comfortable with transgender manager.**

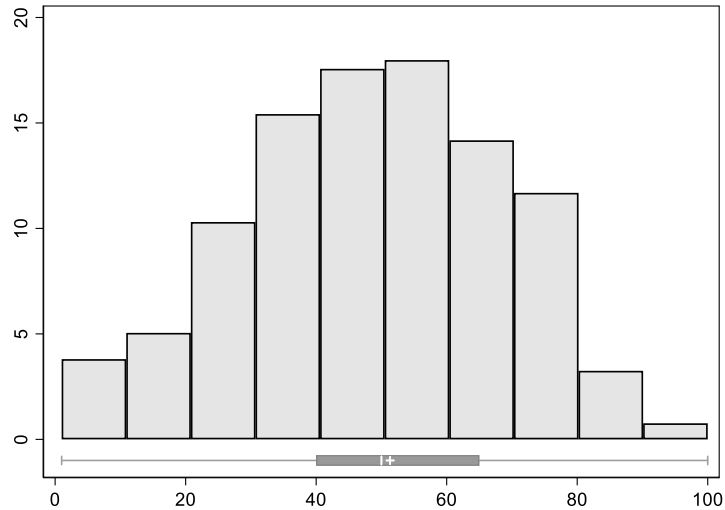

**Panel B: Respondent would not be comfortable with transgender manager.**

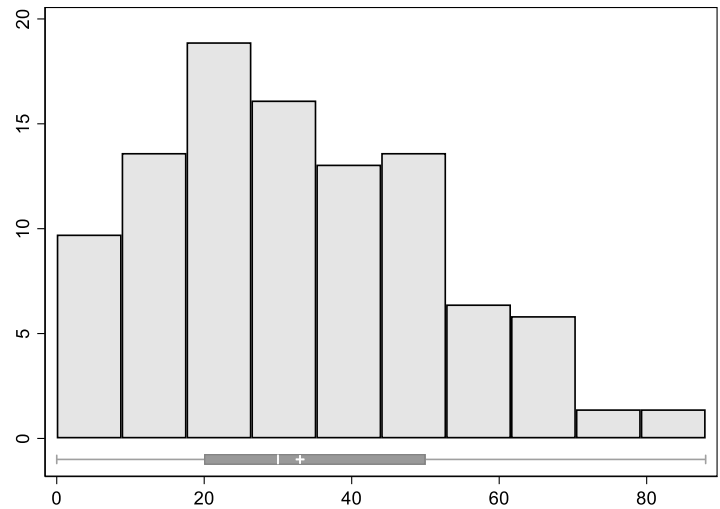

The original survey question is "Out of every 100 people in the general US population, I think approximately __ out of 100 would be comfortable with having a transgender manager at work." The box plot below each histogram reports minimum and maximum values, 25th and 75th percentiles, as well as mean and median. Within each box plot, the white vertical line " | " indicates the median, the white " + " symbol indicates the mean. Number of observations: 1,806. Source: 2022 Prolific List Experiment.

**Figure B5: Respondent thinks X/100 would agree that the law should prohibit employment discrimination against transgender individuals. By responses to direct question.**

**Panel A: Respondent supports transgender employment non-discrimination protection.**

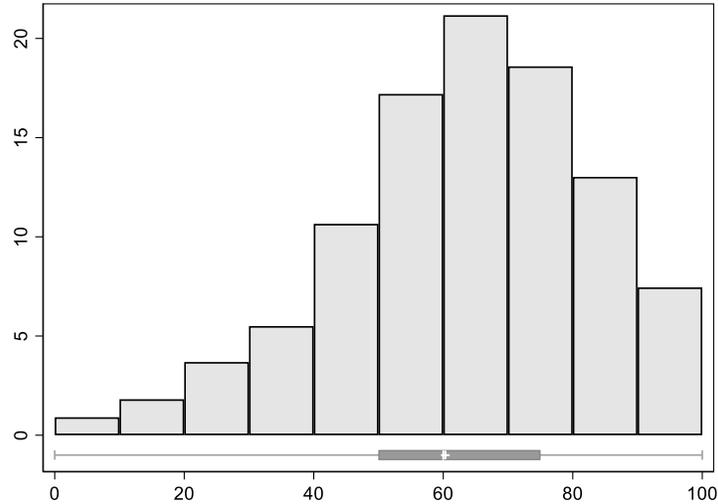

**Panel B: Respondent does not support transgender employment non-discrimination protection.**

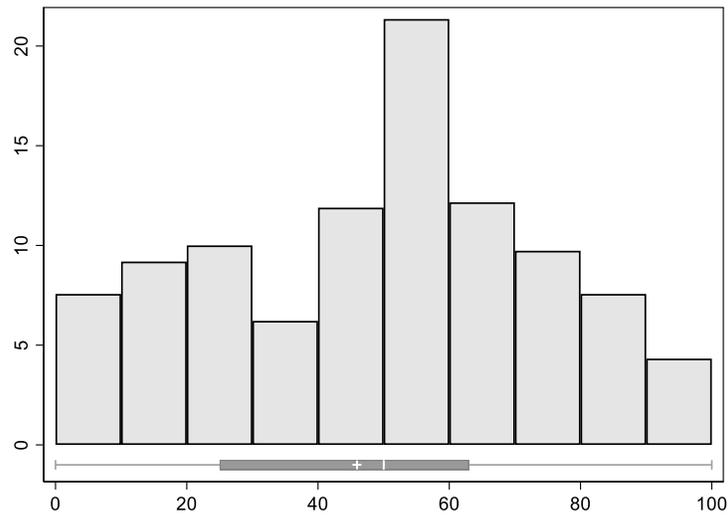

The original survey question is "Out of every 100 people in the general US population, I think approximately __ out of 100 would agree that the law should prohibit employment discrimination against transgender individuals." The box plot below each histogram reports minimum and maximum values, 25[th] and 75[th] percentiles, as well as mean and median. Within each box plot, the white vertical line " | " indicates the median, the white " + " symbol indicates the mean.. Number of observations: 1,806. Source: 2022 Prolific List Experiment.

**Table B1: Sample sizes by gender identity and sex at birth.**

|  | Sex assigned at birth | | | |
|---|---|---|---|---|
|  | Female | Male | Total | |
| Gender identity | (1) | (2) | (3) | |
| Female | 906 | 5 | 911 | 50.70% |
| Male | 7 | 862 | 869 | 48.36% |
| Transgender | 4 | 6 | 10 | 0.56% |
| Non-binary/other | 11 | 9 | 20 | 1.11% |
| Total | 922 | 875 | | |
|  | 51.31% | 48.69% | | |

Original question (columns): "What sex were you assigned at birth, on the original birth certificate?". Original question (rows): "How do you describe yourself? (Check all that apply)". Note that participants could select more than one option for gender. Note that 9 participants (6 female at birth, 3 male at birth) did not select any option for gender. Source: 2022 Prolific List Experiment.

**Table B2: Sample sizes by sexual orientation.**

|  | Non-heterosexual | Heterosexual | Total | |
|---|---|---|---|---|
|  | (1) | (2) | (3) | |
| Gay or Lesbian | 71 | 1 | 72 | 3.99% |
| Straight | 3 | 1,528 | 1,531 | 84.77% |
| Bisexual | 140 | 9 | 149 | 8.25% |
| Something else | 31 | 12 | 43 | 2.38% |
| Don't know | 4 | 7 | 11 | 0.61% |
| Total | 249 | 1,557 | 1,806 | |
|  | 13.79% | 86.21% | | |

Original question (columns): "Are you heterosexual/straight?". Original question (rows): "Which of the following best represents how you think of yourself?". Source: 2022 Prolific List Experiment.

**Table B3: List experiments. Difference-in-means comparisons and robustness checks.**

|  | List A | List B | Double list | Direct question | (1)-(2) | (3)-(4) |
|---|---|---|---|---|---|---|
|  | (1) | (2) | (3) | (4) | (5) | (6) |
| *Panel A: Pooled data* | | | | | | |
| Transgender manager | 0.719 | 0.741 | 0.730 | 0.801 | -0.022 | -0.070*** |
|  | (0.030) | (0.034) | (0.020) | (0.009) | | |
| Trans employment non-discrimination | 0.734 | 0.740 | 0.737 | 0.795 | -0.006 | -0.058*** |
|  | (0.030) | (0.031) | (0.020) | (0.010) | | |
| *Panel B: Excluding pilot data* | | | | | | |
| Transgender manager | 0.723 | 0.738 | 0.731 | 0.801 | -0.016 | -0.071*** |
|  | (0.033) | (0.036) | (0.022) | (0.009) | | |
| Trans employment non-discrimination | 0.743 | 0.749 | 0.746 | 0.795 | -0.006 | -0.049** |
|  | (0.033) | (0.033) | (0.022) | (0.010) | | |
| *Panel C: Adjusted standard errors for stratification* | | | | | | |
| Transgender manager | 0.719 | 0.741 | 0.730 | 0.801 | -0.022 | -0.070*** |
|  | (0.030) | (0.034) | (0.021) | (0.009) | | |
| Trans employment non-discrimination | 0.734 | 0.740 | 0.737 | 0.795 | -0.006 | -0.058*** |
|  | (0.030) | (0.031) | (0.020) | (0.009) | | |
| *Panel D: Excluding too slow and too fast responses* | | | | | | |
| Transgender manager | 0.708 | 0.730 | 0.719 | 0.795 | -0.021 | -0.076*** |
|  | (0.032) | (0.035) | (0.022) | (0.010) | | |
| Trans employment non-discrimination | 0.738 | 0.740 | 0.739 | 0.794 | -0.003 | -0.055*** |
|  | (0.032) | (0.033) | (0.021) | (0.010) | | |

\* $p < 0.10$, ** $p < 0.05$, *** $p < 0.01$. Standard errors in brackets. Standard errors in Panel C have been adjusted for age, sex, and race stratification using the command *svyset* in Stata. In Panel D, we exclude responses (a total of 183 responses) that were submitted too fast (top 5% in terms of speed) or too slow (bottom 5% in terms of speed). Transgender manager key statement: "I would be comfortable having a transgender manager at work". Trans employment non-discrimination key statement: "I think the law should prohibit employment discrimination against transgender individuals". Number of observations: 1,806 (Panels A and C), 1,505 (Panel B), and 1,623 (Panel D). Source: 2022 Prolific List Experiment.

**Table B4: List experiments. Difference-in-means comparisons. Heterogeneity analysis by sex assigned at birth.**

|  | Double list | Direct question | (1)-(2) |
|---|---|---|---|
|  | (1) | (2) | (3) |
| *Panel A: Transgender manager* | | | |
| Women (female at birth), N=928 | 0.801 | 0.852 | -0.052** |
|  | (0.027) | (0.012) |  |
| Men (male at birth), N=878 | 0.658 | 0.746 | -0.088*** |
|  | (0.031) | (0.015) |  |
| Difference between women and men | 0.143*** | 0.106*** | 0.037 |
| *Panel B: Trans employment non-discrimination protection* | | | |
| Women (female at birth), N=928 | 0.768 | 0.843 | -0.075*** |
|  | (0.028) | (0.012) |  |
| Men (male at birth), N=878 | 0.701 | 0.745 | -0.044 |
|  | (0.029) | (0.015) |  |
| Difference between women and men | 0.067* | 0.098*** | -0.031 |

* $p < 0.10$, ** $p < 0.05$, *** $p < 0.01$. Standard errors in brackets. Transgender manager key statement: "I would be comfortable having a transgender manager at work". Trans employment non-discrimination protection key statement: "I think the law should prohibit employment discrimination against transgender individuals". Sex at birth question: "What sex were you assigned at birth, on the original birth certificate?". Number of observations: 1,806. Source: 2022 Prolific List Experiment.

**Table B5: List experiments. Difference-in-means comparisons. Heterogeneity analysis by sexual orientation.**

|  | Double list | Direct question | (1)-(2) |
|---|---|---|---|
|  | (1) | (2) | (3) |
| *Panel A: Transgender manager* | | | |
| Heterosexual, N=1,557 | 0.688 | 0.775 | -0.087*** |
|  | (0.022) | (0.011) |  |
| Non-heterosexual, N=249 | 0.986 | 0.960 | 0.026 |
|  | (0.044) | (0.012) |  |
| Difference by sexual orientation | -0.298*** | -0.185*** | -0.113** |
| *Panel B: Trans employment non-discrimination protection* | | | |
| Heterosexual, N=1,557 | 0.721 | 0.776 | -0.055** |
|  | (0.022) | (0.011) |  |
| Non-heterosexual, N=249 | 0.838 | 0.916 | -0.077* |
|  | (0.042) | (0.018) |  |
| Difference by sexual orientation | -0.117** | -0.140*** | 0.023 |

* $p < 0.10$, ** $p < 0.05$, *** $p < 0.01$. Standard errors in brackets. Transgender manager key statement: "I would be comfortable having a transgender manager at work". Trans employment non-discrimination protection key statement: "I think the law should prohibit employment discrimination against transgender individuals". Sexual orientation question: "Are you heterosexual/straight?". Number of observations: 1,806. Source: 2022 Prolific List Experiment.

**Table B6: List experiments. Difference-in-means comparisons. Heterogeneity analysis by political party affiliation.**

|  | Double list | Direct question | (1)-(2) |
|---|---|---|---|
|  | (1) | (2) | (3) |
| *Panel A: Transgender manager* | | | |
| Democrats, N=873 | 0.879 | 0.930 | -0.052** |
|  | (0.025) | (0.009) |  |
| Republicans, N=350 | 0.424 | 0.509 | -0.084* |
|  | (0.050) | (0.027) |  |
| Independents, N=583 | 0.685 | 0.782 | -0.098*** |
|  | (0.037) | (0.017) |  |
| Difference between Democrats and Republicans | 0.454*** | 0.422*** | 0.033 |
| Difference between Democrats and Independents | 0.194*** | 0.148*** | 0.046 |
| Difference between Republicans and Independents | -0.260*** | -0.274*** | 0.013 |
| *Panel B: Trans employment non-discrimination protection* | | | |
| Democrats, N=873 | 0.875 | 0.899 | -0.024 |
|  | (0.024) | (0.010) |  |
| Republicans, N=350 | 0.481 | 0.534 | -0.053 |
|  | (0.052) | (0.027) |  |
| Independents, N=583 | 0.688 | 0.796 | -0.108*** |
|  | (0.038) | (0.017) |  |
| Difference between Democrats and Republicans | 0.394*** | 0.365*** | 0.029 |
| Difference between Democrats and Independents | 0.187*** | 0.103*** | 0.083* |
| Difference between Republicans and Independents | -0.207*** | -0.262*** | 0.055 |

\* $p < 0.10$, \*\* $p < 0.05$, \*\*\* $p < 0.01$. Standard errors in brackets. Transgender manager key statement: "I would be comfortable having a transgender manager at work". Trans employment non-discrimination protection key statement: "I think the law should prohibit employment discrimination against transgender". Political party affiliation question: "Generally speaking, do you usually think of yourself as a Republican, Democrat, or Independent/Other? Choose the option that best describes you". Number of observations: 1,806. Source: 2022 Prolific List Experiment.

**Table B7: List experiments. Difference-in-means comparisons. Heterogeneity analysis by race.**

|  | Double list | Direct question | (1)-(2) |
|---|---|---|---|
|  | (1) | (2) | (3) |
| *Panel A: Transgender manager* | | | |
| White individuals, N=1,345 | 0.730 | 0.799 | -0.069*** |
|  | (0.023) | (0.011) |  |
| Other or multiple races, N=448 | 0.723 | 0.808 | -0.085** |
|  | (0.042) | (0.019) |  |
| Difference between racial groups | 0.007 | -0.009 | 0.016 |
| *Panel B: Trans employment non-discrimination protection* | | | |
| White individuals, N=1,345 | 0.719 | 0.797 | -0.078*** |
|  | (0.023) | (0.011) |  |
| Other or multiple races, N=448 | 0.797 | 0.797 | 0.000 |
|  | (0.041) | (0.019) |  |
| Difference between racial groups | -0.078 | 0.000 | -0.078* |

* $p < 0.10$, ** $p < 0.05$, *** $p < 0.01$. Standard errors in brackets. Transgender manager key statement: "I would be comfortable having a transgender manager at work". Trans employment non-discrimination protection key statement: "I think the law should prohibit employment discrimination against transgender individuals". Race question" What is your race? Choose all that apply". "Other or multiple races" includes Black or African American, American Indian or Alaskan Native, Asian or Native Hawaiian or Pacific Islander, Some Other Race, and individuals who selected more than one race (including those who selected "white" as one of their race categories). 13 participants who did not select any race have been excluded from this analysis. Number of observations: 1,793. Source: 2022 Prolific List Experiment.

**Table B8: List experiments. Difference-in-means comparisons. Heterogeneity analysis by age.**

|  | Double list | Direct question | (1)-(2) |
|---|---|---|---|
|  | (1) | (2) | (3) |
| *Panel A: Transgender manager* | | | |
| Below median age (18-44), N=913 | 0.795 | 0.843 | -0.048** |
|  | (0.028) | (0.012) |  |
| Above median age (45 or older), N=893 | 0.654 | 0.757 | -0.103*** |
|  | (0.030) | (0.014) |  |
| Difference between younger and older group | 0.141*** | 0.086*** | 0.055 |
| *Panel B: Trans employment non-discrimination protection* | | | |
| Below median age (18-44), N=913 | 0.794 | 0.841 | -0.048* |
|  | (0.026) | (0.012) |  |
| Above median age (45 or older), N=893 | 0.684 | 0.748 | -0.064** |
|  | (0.031) | (0.015) |  |
| Difference between younger and older group | 0.110*** | 0.093*** | 0.017 |

\* $p < 0.10$, \*\* $p < 0.05$, \*\*\* $p < 0.01$. Standard errors in brackets. Transgender manager key statement: "I would be comfortable having a transgender manager at work". Trans employment non-discrimination protection key statement: "I think the law should prohibit employment discrimination against transgender individuals". Age question: "What is your age in years?" Number of observations: 1,806. Source: 2022 Prolific List Experiment.

**Table B9: List experiments. Difference-in-means comparisons. Heterogeneity analysis by sexual attraction.**

|  | Double list | Direct question | (1)-(2) |
|---|---|---|---|
|  | (1) | (2) | (3) |
| *Panel A: Transgender manager* | | | |
| Exclusively attracted to a different sex, N= 1,328 | 0.668 | 0.752 | -0.084*** |
|  | (0.025) | (0.012) |  |
| Other, N= 478 | 0.908 | 0.935 | -0.028 |
|  | (0.034) | (0.011) |  |
| Difference by sexual attraction | -0.239*** | -0.183*** | -0.057 |
| *Panel B: Trans employment non-discrimination protection* | | | |
| Exclusively attracted to a different sex, N=1,328 | 0.703 | 0.758 | -0.055** |
|  | (0.025) | (0.012) |  |
| Other, N=478 | 0.829 | 0.900 | -0.071** |
|  | (0.033) | (0.014) |  |
| Difference by sexual attraction | -0.126*** | -0.142*** | 0.016 |

\* $p < 0.10$, ** $p < 0.05$, *** $p < 0.01$. Standard errors in brackets. Transgender manager key statement: "I would be comfortable having a transgender manager at work". Trans employment non-discrimination protection key statement: "I think the law should prohibit employment discrimination against transgender individuals". The sexual attraction category "Other" includes participants attracted to both females and males, participants attracted to same-sex individuals (same-sex based on sex at birth), and participants who selected the option "Other" when asked about their sexual attraction. Number of observations: 1,806. Source: 2022 Prolific List Experiment.

**Table B10: List experiments. Difference-in-means comparisons. Heterogeneity analysis by income.**

|  | Double list | Direct question | (1)-(2) |
|---|---|---|---|
|  | (1) | (2) | (3) |
| *Panel A: Transgender manager* | | | |
| Below median income (<$60,000), N=862 | 0.722 | 0.796 | -0.074*** |
|  | (0.030) | (0.014) |  |
| Above median income (≥$60,000), N=944 | 0.738 | 0.805 | -0.067*** |
|  | (0.028) | (0.013) |  |
| Difference between below and above $60,000 | -0.016 | -0.009 | -0.007 |
| *Panel B: Trans employment non-discrimination protection* | | | |
| Below median income (<$60,000), N=862 | 0.754 | 0.794 | -0.039 |
|  | (0.030) | (0.014) |  |
| Above median income (≥$60,000), N=944 | 0.721 | 0.797 | -0.075*** |
|  | (0.027) | (0.013) |  |
| Difference between below and above $60,000 | 0.033 | -0.003 | 0.036 |

\* $p < 0.10$, ** $p < 0.05$, *** $p < 0.01$. Standard errors in brackets. Transgender manager key statement: "I would be comfortable having a transgender manager at work". Trans employment non-discrimination protection key statement: "I think the law should prohibit employment discrimination against transgender individuals". Income question: "What is your household income before taxes?" Number of observations: 1,806. Source: 2022 Prolific List Experiment.

**Table B11: List experiments. Difference-in-means comparisons. Heterogeneity analysis by education.**

|  | Double list | Direct question | (1)-(2) |
|---|---|---|---|
|  | (1) | (2) | (3) |
| *Panel A: Transgender manager* | | | |
| Less than a Bachelor's degree, N=753 | 0.690 | 0.776 | -0.086*** |
|  | (0.033) | (0.015) |  |
| Bachelor's degree or higher, N=1,053 | 0.759 | 0.819 | -0.060** |
|  | (0.026) | (0.012) |  |
| Difference between education groups | -0.069* | -0.043** | -0.026 |
| *Panel B: Trans employment non-discrimination protection* | | | |
| Less than a Bachelor's degree, N=753 | 0.741 | 0.786 | -0.045 |
|  | (0.032) | (0.015) |  |
| Bachelor's degree or higher, N=1,053 | 0.735 | 0.802 | -0.067*** |
|  | (0.026) | (0.012) |  |
| Difference between education groups | 0.006 | -0.015 | 0.021 |

* $p < 0.10$, ** $p < 0.05$, *** $p < 0.01$. Standard errors in brackets. Transgender manager key statement: "I would be comfortable having a transgender manager at work". Trans employment non-discrimination protection key statement: "I think the law should prohibit employment discrimination against transgender individuals". Education question: "What is the highest level of education you've completed? (choose one) (If currently enrolled, mark the previous grade or highest degree received.)" Number of observations: 1,806. Source: 2022 Prolific List Experiment.

**Table B12: List experiments. Difference-in-means comparisons. Heterogeneity analysis by employment status.**

|  | Double list | Direct question | (1)-(2) |
|---|---|---|---|
|  | (1) | (2) | (3) |
| *Panel A: Transgender manager* | | | |
| Employed or self-employed, N=1,210 | 0.715 | 0.794 | -0.079*** |
|  | (0.025) | (0.012) |  |
| Unemployed or not in the labor force, N=596 | 0.758 | 0.814 | -0.056* |
|  | (0.037) | (0.016) |  |
| Difference between employment groups | -0.043 | -0.020 | -0.024 |
| *Panel B: Trans employment non-discrimination protection* | | | |
| Employed or self-employed, N=1,210 | 0.734 | 0.793 | -0.059** |
|  | (0.024) | (0.012) |  |
| Unemployed or not in the labor force, N=596 | 0.741 | 0.799 | -0.057 |
|  | (0.036) | (0.016) |  |
| Difference between employment groups | -0.007 | -0.005 | -0.002 |

\* $p < 0.10$, \*\* $p < 0.05$, \*\*\* $p < 0.01$. Standard errors in brackets. Transgender manager key statement: "I would be comfortable having a transgender manager at work". Trans employment non-discrimination protection key statement: "I think the law should prohibit employment discrimination against transgender individuals". "Unemployed or not in the labor force" includes homemakers, students, retired individuals, individuals unable to work, and individuals out of work. Number of observations: 1,806. Source: 2022 Prolific List Experiment.

**Table B13: List experiments. Difference-in-means comparisons. Heterogeneity analysis by managerial experience.**

|  | Double list | Direct question | (1)-(2) |
|---|---|---|---|
|  | (1) | (2) | (3) |
| *Panel A: Transgender manager* | | | |
| Has managerial experience, N=983 | 0.690 | 0.784 | -0.095*** |
|  | (0.028) | (0.013) | |
| No managerial experience, N=749 | 0.787 | 0.821 | -0.034 |
|  | (0.031) | (0.014) | |
| Difference by managerial experience | -0.098** | -0.037* | -0.061 |
| *Panel B: Trans employment non-discrimination protection* | | | |
| Has managerial experience, N=983 | 0.705 | 0.774 | -0.069** |
|  | (0.028) | (0.013) | |
| No managerial experience, N=749 | 0.775 | 0.829 | -0.054* |
|  | (0.031) | (0.014) | |
| Difference by managerial experience | -0.070* | -0.055*** | -0.015 |

* $p < 0.10$, ** $p < 0.05$, *** $p < 0.01$. Standard errors in brackets. Transgender manager key statement: "I would be comfortable having a transgender manager at work". Trans employment non-discrimination protection key statement: "I think the law should prohibit employment discrimination against transgender individuals". Managerial experience question (collected by Prolific): "Do you have any experience being in a management position?" Number of observations: 1,732 (74 missing values are excluded). Source: 2022 Prolific List Experiment.

**Table B14: List experiments. Difference-in-means comparisons. Heterogeneity analysis by current religious affiliation.**

|  | Double list | Direct question | (1)-(2) |
|---|---|---|---|
|  | (1) | (2) | (3) |
| *Panel A: Transgender manager* | | | |
| Christian (any denomination), N=826 | 0.622 | 0.706 | -0.084*** |
|  | (0.032) | (0.016) |  |
| Not religious, N=836 | 0.824 | 0.891 | -0.067*** |
|  | (0.027) | (0.011) |  |
| Difference by current religious affiliations | -0.202*** | -0.185*** | -0.017 |
| *Panel B: Trans employment non-discrimination protection* | | | |
| Christian (any denomination), N=826 | 0.648 | 0.717 | -0.068** |
|  | (0.032) | (0.016) |  |
| Not religious, N=836 | 0.829 | 0.870 | -0.041 |
|  | (0.026) | (0.012) |  |
| Difference by current religious affiliations | -0.180*** | -0.153*** | -0.027 |

\* p < 0.10, \*\* p < 0.05, \*\*\* p < 0.01. Standard errors in brackets. Transgender manager key statement: "I would be comfortable having a transgender manager at work". Trans employment non-discrimination protection key statement: "I think the law should prohibit employment discrimination against transgender individuals". Religion question: "What is your current religious affiliation?" Number of observations: 1,662 (144 participants with other religious affiliations excluded from this comparison). Source: 2022 Prolific List Experiment.

**Table B15: List experiments. Difference-in-means comparisons. Heterogeneity analysis by religion importance in participant's life.**

|  | Double list | Direct question | (1)-(2) |
|---|---|---|---|
|  | (1) | (2) | (3) |
| *Panel A: Transgender manager* | | | |
| Religion important in life, N=754 | 0.627 | 0.686 | -0.059* |
|  | (0.035) | (0.017) |  |
| Religion not important in life, N=1,052 | 0.805 | 0.883 | -0.078*** |
|  | (0.025) | (0.010) |  |
| Difference by religion importance | -0.178*** | -0.197*** | 0.020 |
| *Panel B: Trans employment non-discrimination protection* | | | |
| Religion important in life, N=754 | 0.613 | 0.704 | -0.091*** |
|  | (0.033) | (0.017) |  |
| Religion not important in life, N=1,052 | 0.824 | 0.860 | -0.036 |
|  | (0.024) | (0.011) |  |
| Difference by religion importance | -0.211*** | -0.156*** | -0.055 |

* $p < 0.10$, ** $p < 0.05$, *** $p < 0.01$. Standard errors in brackets. Transgender manager key statement: "I would be comfortable having a transgender manager at work". Trans employment non-discrimination protection key statement: "I think the law should prohibit employment discrimination against transgender individuals". Religiosity question: "How important is religion in your life?" Participants who answered "Very Important" or "Somewhat important" coded as "Religion important in life". Participants who answered "Not too important" or "Not at all important" coded as "Religion not important in life". Number of observations: 1,806. Source: 2022 Prolific List Experiment.

**Table B16: List experiments. Difference-in-means comparisons. Heterogeneity analysis by geographic location.**

|  | Double list | Direct question | (1)-(2) |
|---|---|---|---|
|  | (1) | (2) | (3) |
| *Panel A: Transgender manager* | | | |
| North-East, N=381 | 0.787 | 0.877 | -0.090** |
|  | (0.044) | (0.017) |  |
| Midwest, N=389 | 0.754 | 0.820 | -0.066* |
|  | (0.042) | (0.020) |  |
| West, N=271 | 0.706 | 0.786 | -0.080* |
|  | (0.050) | (0.025) |  |
| South, N=765 | 0.698 | 0.758 | -0.060** |
|  | (0.033) | (0.015) |  |
| *Panel B: Trans employment non-discrimination protection* | | | |
| North-East, N=381 | 0.758 | 0.866 | -0.108** |
|  | (0.041) | (0.017) |  |
| Midwest, N=389 | 0.735 | 0.789 | -0.054 |
|  | (0.044) | (0.021) |  |
| West, N=271 | 0.801 | 0.790 | 0.011 |
|  | (0.051) | (0.025) |  |
| South, N=765 | 0.704 | 0.765 | -0.061* |
|  | (0.032) | (0.015) |  |

\* $p < 0.10$, ** $p < 0.05$, *** $p < 0.01$. Standard errors in brackets. Transgender manager key statement: "I would be comfortable having a transgender manager at work". Trans employment non-discrimination protection key statement: "I think the law should prohibit employment discrimination against transgender individuals". Participants are divided in groups based on the US state where they lived at the time of the survey. Number of observations: 1,806. Source: 2022 Prolific List Experiment.

**Table B17: List experiments. Difference-in-means comparisons. Heterogeneity analysis by beliefs regarding attitudes of the general U.S. population.**

|  | Double list | Direct question | (1)-(2) |
|---|---|---|---|
|  | (1) | (2) | (3) |
| *Panel A: Transgender manager* | | | |
| Respondents believe 50% or more of Americans would be comfortable with a transgender manager at work, N=963 | 0.837 | 0.899 | -0.063** |
|  | (0.027) | (0.010) |  |
| Respondents believe less than 50% of Americans would be comfortable with a transgender manager at work, N=843 | 0.610 | 0.688 | -0.078*** |
|  | (0.030) | (0.016) |  |
| Difference by beliefs | 0.226*** | 0.211*** | 0.015 |
| *Panel B: Trans employment non-discrimination protection* | | | |
| Respondents believe 50% or more of Americans would agree that the law should prohibit employment discrimination against transgender individuals, N=1,316 | 0.765 | 0.845 | -0.080*** |
|  | (0.023) | (0.010) |  |
| Respondents believe less than 50% of Americans would agree that the law should prohibit employment discrimination against transgender individuals, N=490 | 0.661 | 0.661 | 0.000 |
|  | (0.040) | (0.021) |  |
| Difference by beliefs | 0.104** | 0.184*** | -0.080* |

\* $p < 0.10$, \*\* $p < 0.05$, \*\*\* $p < 0.01$. Standard errors in brackets. Transgender manager key statement: "I would be comfortable having a transgender manager at work". Trans employment non-discrimination protection key statement: "I think the law should prohibit employment discrimination against transgender individuals". Number of observations: 1,806. Source: 2022 Prolific List Experiment.

**Table B18: Balance table.**

| Variables | List A with key statement (List B without key statement) | List A without key statement (List B with key statement) | Difference |
|---|---|---|---|
| Age: between 18 and 34 | 0.316 | 0.352 | -0.036 |
| Age: between 35 and 49 | 0.255 | 0.252 | 0.003 |
| Age: between 50 and 64 | 0.286 | 0.278 | 0.008 |
| Age: 65 or over | 0.142 | 0.117 | 0.025 |
| Sex assigned at birth: Female | 0.534 | 0.494 | 0.040* |
| Race: White only | 0.752 | 0.737 | 0.015 |
| Race: Black or African American only | 0.145 | 0.125 | 0.021 |
| Race: Asian or Native Hawaiian or Pacific Islander only | 0.062 | 0.069 | -0.006 |
| Ethnicity: Hispanic | 0.046 | 0.062 | -0.016 |
| Marital status: Now married | 0.454 | 0.428 | 0.026 |
| Marital status: Widowed | 0.024 | 0.030 | -0.005 |
| Marital status: Separated | 0.008 | 0.012 | -0.004 |
| Marital status: Divorced | 0.127 | 0.115 | 0.012 |
| Marital status: Never married | 0.387 | 0.415 | -0.028 |
| Education: High school, GED, or less | 0.104 | 0.109 | -0.005 |
| Education: Some college credits, no degree | 0.205 | 0.194 | 0.011 |
| Education: Associate's degree | 0.115 | 0.105 | 0.010 |
| Education: Bachelor's degree or equivalent | 0.390 | 0.383 | 0.006 |
| Education: Master's degree or higher | 0.185 | 0.208 | -0.022 |
| Employment: Employed for wages | 0.542 | 0.529 | 0.012 |
| Employment: Self-employed | 0.144 | 0.125 | 0.019 |
| Employment: Out of work for 1 year or more | 0.052 | 0.063 | -0.011 |
| Employment: Out of work for less than 1 year | 0.011 | 0.015 | -0.004 |
| Employment: Homemaker | 0.058 | 0.042 | 0.016 |
| Employment: Student | 0.047 | 0.062 | -0.015 |
| Employment: Retired | 0.125 | 0.133 | -0.007 |
| Employment: Unable to work | 0.021 | 0.031 | -0.010 |
| Household income: less than $60,000 | 0.481 | 0.474 | 0.007 |
| Political party affiliation: Democrat | 0.484 | 0.483 | 0.001 |
| Political party affiliation: Republican | 0.201 | 0.187 | 0.014 |
| Political party affiliation: Independent | 0.315 | 0.330 | -0.015 |
| Urbanicity: Rural area | 0.115 | 0.136 | -0.020 |
| Urbanicity: Small city or town | 0.296 | 0.285 | 0.011 |
| Urbanicity: Suburb near a large city | 0.361 | 0.335 | 0.026 |
| Urbanicity: Large city | 0.228 | 0.244 | -0.017 |
| Currently live in: North-East | 0.208 | 0.214 | -0.007 |
| Currently live in: Midwest | 0.201 | 0.230 | -0.029 |
| Currently live in: South | 0.437 | 0.410 | 0.027 |
| Currently live in: West | 0.154 | 0.146 | 0.008 |
| Observations | 901 | 905 | |

Source: 2022 Prolific List Experiment. * $p < 0.10$, ** $p < 0.05$, *** $p < 0.01$.

**Appendix C. List experiment and survey instructions**

### Prolific ID Entry

Thank you for participating in this survey. Before we begin, please fill out the field below.

Paste your Prolific ID here

[                    ]

[ Next ]

# Consent Page

Please review the information contained in the form below.

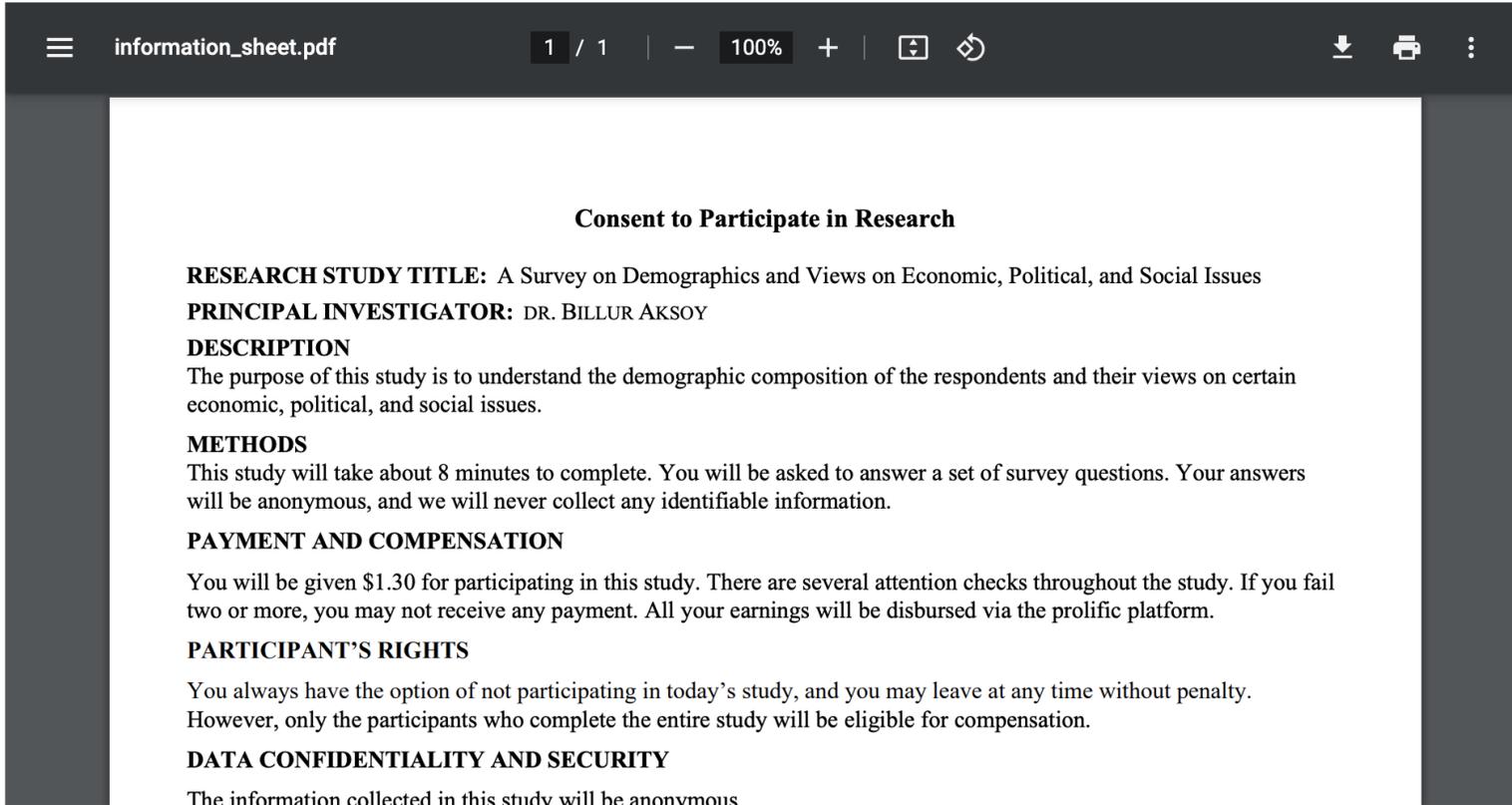

If you do not consent to participate in this study, please go back to Prolific now and mark this study as Incomplete. By clicking below, you confirm that you are at least 18 years of age and that you consent to participation in the research study described above.

Next

# Overview of Study

## Welcome!

In this study, you will be asked to answer some questions. Please try to answer these questions truthfully. Your answers will be <u>anonymous</u>.

The study will take about 8 minutes.

## How much payment will I receive for my participation?

You will be paid 1.30 USD for completing the study.

## Attention Checks

There will be several **<u>Attention Check</u>** questions throughout this study meant to test whether you are paying attention. If you fail to correctly complete two or more of these Attention Check questions, you may not be paid.

Next

## Instructions

In the following pages, you will be presented with lists of statements that may or may not be true for you. The statements will be about yourself and your views on social issues. We would like to know how many of the statements within each list are true for you. In these lists, we are **not** asking which specific statements are true for you, we are only asking **how many** of them are true for you.

On the following page, we will give you an example. Please click next when you are ready.

Next

## Example:

*We want to be sure that you understand how this works. Here is an example:*

> There are three statements in the list below. We would like to know how many of them are true for you.
>
> - I own an orange t-shirt.
> - My household has at least four pets.
> - I regularly recycle.
>
> Please enter the total number of the above statements that are true for you:
>
> [    ]

*Suppose that you do own an orange t-shirt and your household has at least four pets. But you do not recycle. In that case, two of the above statements are true for you. Hence, you would indicate this by entering 2 in the answer box.*

**Please Note:** We are not asking which specific statements in these lists are true for you. We are only asking how many of them are true for you.

Please click NEXT when you are ready.

[Next]

If you would like to review the instructions one more time, please Click Here

In the following pages, you will be presented with lists of statements that may or may not be true for you. The statements will be about yourself and your views on social issues. We would like to know how many of the statements within each list are true for you. In these lists, we are not asking which specific statements are true for you, we are only asking how many of them are true for you.

When you are ready, please click next to start.

Next

**{Manager List A}**

There are four statements in the list below. How many of the following statements are true for you?

- I support the Black Lives Matter movement.
- I can fluently speak at least three languages.
- I have a driver's license.
- I think COVID-19 health risks were overstated.

Please enter the total number of the above statements that are true for you:

[ ]

Next

**{Manager List B}**

There are five statements in the list below. How many of the following statements are true for you?

- I think parents should be able to opt their children out of a COVID-19 school mask mandate for any reason.
- I would be comfortable having a transgender manager at work.
- I own a car.
- I have visited more than twenty countries.
- I think the law should prohibit employment discrimination against African Americans.

Please enter the total number of the above statements that are true for you:

[ ]

[Next]

**{Employ Non-Discrim List A}**

There are four statements in the list below. How many of the following statements are true for you?

- I would vote for a political candidate who is pro-life (anti-abortion).
- I have personally met the current Pope.
- I think recreational marijuana use should be legal.
- I own a smartphone.

Please enter the total number of the above statements that are true for you:

[          ]

[Next]

**{Employ Non-Discrim List B}**

There are five statements in the list below. How many of the following statements are true for you?

- I would vote for a political candidate who is pro-choice (supports abortion rights).
- I think gun control laws should be relaxed.
- I have at least one social media account (e.g., Facebook, Twitter, Instagram).
- I have personally met the current U.S. President.
- I think the law should prohibit employment discrimination against transgender individuals.

Please enter the total number of the above statements that are true for you:

[        ]

[Next]

There are five statements in the list below. How many of the following statements are true for you?

- I usually respond to my emails within 24 hours.
- I am concerned that the media in the United States is biased.
- Please put seven as your answer below regardless of how many of the others are true for you.
- This is because we would like to see whether you are reading each item carefully.
- Again, please put seven for your answer below.

Please enter the total number of the above statements that are true for you:

Next

---------- ---------- ---------- ---------- ---------- ---------- ---------- ---------- ---------- ---------- ---------- ----------

## Survey

Next, we will ask you some demographics questions about yourself as well as your opinion on certain issues. Please answer the following questions to the best of your ability. Again, please remember that your answers will be completely <u>anonymous</u>.

Next

# Survey

What is your race? Choose all that apply.

- [ ] White
- [ ] Black or African American
- [ ] American Indian or Alaskan Native
- [ ] Asian or Native Hawaiian or Pacific Islander
- [ ] Some Other Race

Are you of Hispanic, Latino, or Spanish origin?

- ○ Yes
- ○ No

What is your age in years?

[          ]

What is your marital status? (choose one)

- ○ Now married
- ○ Widowed
- ○ Divorced
- ○ Separated
- ○ Never married

[Next]

# Survey

How many people live in your household including yourself?

How many children less than 18 years of age live in your household? If none, please put 0. Number of children:

What is the highest level of education you've completed? (choose one) (If currently enrolled, mark the previous grade or highest degree received.)

- ○ High school, GED, or less
- ○ Some college credits, no degree
- ○ Associate's degree (for example: AA, AS)
- ○ Bachelor's degree or equivalent (for example: BA, BS)
- ○ Master's degree or higher (for example: MA, MS, MEng, MEd, MSW, MBA, MD, DDS, DVM, LLB, JD, PhD, EdD)

Are you currently...?

- ○ Employed for wages
- ○ Self-employed
- ○ Out of work for 1 year or more
- ○ Out of work for less than 1 year
- ○ A homemaker
- ○ A student
- ○ Retired
- ○ Unable to work

Next

# Survey

What type of community do you live in?

- ○ Rural area
- ○ Large city
- ○ Small city or town
- ○ Suburb near a large city

In which US state/territory do you currently live?

In which US state/territory did you spend the most time for the first 18 years of your life?

Next

# Survey

Would you be comfortable having a transgender manager at work?

○ Yes
○ No

Do you think the law should prohibit employment discrimination against transgender individuals?

○ Yes
○ No

[Next]

# Survey

Do you think the law should prohibit housing discrimination against transgender individuals?

○ Yes
○ No

Would you be comfortable having a transgender person as a neighbor?

○ Yes
○ No

[Next]

# Survey

Would you be comfortable having an openly lesbian, gay, or bisexual manager at work?

- ○ Yes
- ○ No

Do you think the law should prohibit employment discrimination against lesbian, gay, or bisexual individuals?

- ○ Yes
- ○ No

Do you think the law should prohibit housing discrimination against lesbian, gay, or bisexual individuals?

- ○ Yes
- ○ No

Would you be comfortable having an openly lesbian, gay, or bisexual person as a neighbor?

- ○ Yes
- ○ No

[Next]

# Survey

Do you think that private businesses (such as cake decorators or florists) should be able to refuse service to same-sex couples or other LGBTQ+ individuals for religious reasons?

○ Yes
○ No

Next

# Survey

What sex were you assigned at birth, on your original birth certificate? (choose one)

○ Male

○ Female

How do you describe yourself? (check all that apply)

☐ Male

☐ Female

☐ Transgender

☐ Non-Binary/Other

Are you heterosexual/straight?

○ Yes

○ No

Which of the following best represents how you think of yourself?

○ Gay or Lesbian

○ Straight, that is, not gay or lesbian

○ Bisexual

○ Something else

○ I don't know the answer

Since age 18, have you had at least one same-sex sexual partner?

○ Yes

○ No

People are different in their sexual attraction to other people. Which category below best describes your feelings?

- ○ Only attracted to females
- ○ Mostly attracted to females
- ○ Equally attracted to females and males
- ○ Mostly attracted to males
- ○ Only attracted to males
- ○ Other (please specify below)

[                    ]

Before providing an answer, one should always read the text carefully. To check whether you have been reading the text carefully, we ask you to select the third option below as your answer.

- ○ First
- ○ Second
- ○ Third
- ○ Fourth

[Next]

# Survey

What is your current religious affiliation?

- ○ Christian (any denomination)
- ○ Jewish
- ○ Muslim (any denomination)
- ○ Hindu
- ○ Buddhist
- ○ Asian Folk Religion (e.g., Taoist, Confucian)
- ○ I am not religious
- ○ Some other religious affiliation (please specify below)

[                                    ]

Which of the following religious affiliations best describes how you were raised?

- ○ Christian (any denomination)
- ○ Jewish
- ○ Muslim (any denomination)
- ○ Hindu
- ○ Buddhist
- ○ Asian Folk Religion (e.g., Taoist, Confucian)
- ○ I was not raised in any religion
- ○ Some other religious affiliation (please specify below)

[                                    ]

How important is religion in your life?

- ○ Very important
- ○ Somewhat important
- ○ Not too important
- ○ Not at all important

Generally speaking, do you usually think of yourself as a Republican, Democrat, or Independent/Other? Choose the option that best describes you.

○ Republican
○ Democrat
○ Independent or Other

On a scale of 1-7, 1 being extremely liberal and 7 being extremely conservative, how liberal/conservative would you say your political views on social issues are?

○ 1. Extremely liberal
○ 2. Liberal
○ 3. Slightly liberal
○ 4. Moderate, middle of the road
○ 5. Slightly conservative
○ 6. Conservative
○ 7. Extremely conservative

We would like to be sure that you are reading these questions and not making random decisions. Thus, please select the last option for this question.

○ First
○ Second
○ Last

Who did you vote for in the 2016 presidential election?

○ Donald Trump
○ Hillary Clinton
○ Other
○ Did not vote
○ Not eligible to vote
○ I do not remember

Who did you vote for in the 2020 presidential election?
- ○ Joe Biden
- ○ Donald Trump
- ○ Other
- ○ Did not vote
- ○ Not eligible to vote
- ○ I do not remember

What is your household income before taxes?
- ○ Less than $20,000
- ○ $20,000 - $39,999
- ○ $40,000 - $59,999
- ○ $60,000 - $79,999
- ○ $80,000 - $99,999
- ○ $100,000 - $149,999
- ○ $150,000 - $199,999
- ○ $200,000 or higher

Next

# Survey

Based on your understanding, federal law prohibits employment discrimination on the basis of which of the following characteristics? (check all that apply)

- [ ] Race
- [ ] Disability
- [ ] Sexual orientation
- [ ] Sex
- [ ] Eye color
- [ ] Political beliefs

Next

# Survey

In this part of our survey, we want to know what you think about public perceptions on certain issues in the U.S. When you answer the following questions, please think about the **general U.S. population**.

"Out of every 100 people in the general US population, I think approximately [   ] out of 100 would be comfortable with having a transgender manager at work."

"Out of every 100 people in the general US population, I think approximately [   ] out of 100 would agree that the law should prohibit employment discrimination against transgender individuals."

Next

---------- ---------- ---------- ---------- ---------- ---------- ---------- ---------- ---------- ---------- ----------

Finally, please answer the following question.

Is there anything else you would like share with the researchers?

Next

## Prolific ID Entry

Thank you for participating in this survey. We will process your payment shortly.

**Please enter your Prolific ID again.**

Next

---------- ---------- ---------- ---------- ---------- ---------- ---------- ---------- ---------- ---------- ----------

Thank you for participating in this study. Please follow the link below to return to Prolific.

[Return to Prolific](#)

**List Experiment Instructions Used in First Wave:**

### Instructions

In the following pages, you will be presented with lists of statements that may or may not be true for you. The statements will be about yourself and your views on social issues. And we will ask you **how many** of those statements in each list are true for you.

**Please note that there is no way for us to know which specific statements in these lists are true for you: we will only know how many of them are true for you.**

On the following page, we will give you an example. Please click next when you are ready.

[Next]

## Example:

*We want to be sure that you understand how this works. Here is an example:*

> There are three statements in the list below. We would like to know how many of them are true for you.
>
> - I own an orange t-shirt.
> - My household has at least four pets.
> - I regularly recycle.
>
> Please enter the total number of the above statements that are true for you:

*Suppose that you do own an orange t-shirt and your household has at least four pets. But you do not recycle. In that case, two of the above statements are true for you. Hence, you would indicate this by entering 2 in the answer box.*

**Please Note:** There is no way for us to know which specific statements in these lists are true for you. We will only know how many of them are true for you.

Please click NEXT when you are ready.

Next

If you would like to review the instructions one more time, please [Click Here]

> In the following pages, you will be presented with lists of statements that may or may not be true for you. The statements will be about yourself and your views on social issues. We would like to know how many of them are true for you. We do not want to know which specific ones are true for you. Instead, we just want to know how many of the statements are true for you.

When you are ready, please click next to start.

[Next]